\documentclass[aps,prmaterials,reprint,superscriptaddress,longbibliography,twocolumn]{revtex4-2}
\usepackage[english]{babel}  
\usepackage{comment} 
\usepackage{graphicx} 
\usepackage{amsmath,amssymb} 
\usepackage{bm} 
\usepackage{xcolor} 
\definecolor{lightgray}{gray}{0.6}
\definecolor{medgray}{gray}{0.4}

\usepackage{enumitem}
\setlist{noitemsep,leftmargin=*,topsep=0pt,parsep=0pt}

\newcommand{\mytitle}{Critical nematic correlations throughout the superconducting doping range in \BSCO}

\newcommand {\BSCO}{Bi$_{2-z}$Pb$_z$Sr$_{2-y}$La$_y$CuO$_{6+x}$}
\renewcommand{\vec}[1]{\bm{#1}}

\begin{document}

\title{\mytitle}

\author{Can-Li Song}
\affiliation{Department of Physics, Harvard University, Cambridge, Massachusetts 02138, USA.}
\author{Elizabeth J. Main}
\affiliation{Department of Physics, Harvard University, Cambridge, Massachusetts 02138, USA.}
\author{Forrest Simmons}
\affiliation{Department of Physics and Astronomy, Purdue University, West Lafayette, IN 47907, USA.}
\affiliation{Purdue Quantum Science and Engineering Institute, West Lafayette, IN 47907, USA}
\author{Shuo Liu}
\affiliation{Department of Physics, Purdue University, West Lafayette, IN 47907, USA.}
\author{Benjamin Phillabaum}
\affiliation{Department of Physics, Purdue University, West Lafayette, IN 47907, USA.}
\author{Karin A. Dahmen}
\affiliation{Department of Physics, University of Illinois, Urbana-Champaign, Ilinois 61801, USA.}
\author{E. W. Hudson}
\affiliation{Department of Physics, The Pennsylvania State University, University Park, Pennsylvania 16802, USA.}
\author{Jennifer E. Hoffman}
\email{jhoffman@physics.harvard.edu}
\affiliation{Department of Physics, Harvard University, Cambridge, Massachusetts 02138, USA.}
\author{E. W. Carlson}
\email{ewcarlson@physics.purdue.edu}
\affiliation{Department of Physics, Purdue University, West Lafayette, IN 47907, USA.}
\affiliation{Purdue Quantum Science and Engineering Institute, West Lafayette, IN 47907, USA}

\date{\today}

\begin{abstract}
Charge modulations have been widely observed in cuprates, suggesting their centrality for understanding the high-$T_c$ superconductivity in these materials. However, the dimensionality of these modulations remains controversial, including whether their wavevector is unidirectional or bidirectional, and also whether they extend seamlessly from the surface of the material into the bulk. Material disorder presents severe challenges to understanding the charge modulations through bulk scattering techniques.  We use a local technique, scanning tunneling microscopy, to image the static charge modulations on \BSCO. By comparing the phase correlation length $\xi_{\mathrm{CDW}}$ with the orientation correlation length $\xi_{\mathrm{orient}}$, we show that the charge modulations are more consistent with an underlying unidirectional wave vector. 
By computing new critical exponents at free surfaces including that of the pair connectivity correlation function, we show that these locally 1D charge modulations are actually a bulk effect resulting from 3D criticality throughout the entire superconducting doping range. 
\end{abstract}

\maketitle


Charge order, long seen on the surface of BSCCO  \cite{HoffmanScience2002vortex,HowaldPNAS2003,KohsakaScience2007}, BSCO \cite{WiseNatPhys2008}, and Na-CCOC \cite{HanaguriNature2004,KohsakaScience2007}, has recently been demonstrated in the bulk of many superconducting cuprates by NMR and scattering techniques \cite{WuNature2011,ChangNatPhys2012,GhiringhelliScience2012,CominScience2014,daSilvaNetoScience2015,croft_charge_2014,tabis_charge_2014,da_silva_neto_charge_2015,peng_direct_2016,chang_direct_2012,kang_evolution_2019,ghiringhelli_long-range_2012,WuNature2011,li_multiorbital_2020,abbamonte_spatially_2005,comin_symmetry_2015,da_silva_neto_ubiquitous_2014}. Its apparent universality prioritizes its microscopic understanding and the question of its relationship to superconductivity. However, severe material disorder presents both a challenge and an opportunity \cite{PhillabaumNatComm2012}. The challenge is that material disorder disrupts long-range order and limits macroscopic experimental probes to reporting spatially averaged properties. In particular, while numerous theories rest upon the 1D (stripe) or 2D (checker) nature of the charger order (CO), bulk probes may collect signal from multiple domains, obscuring the underlying dimensionality within a single domain.

The opportunity is for local probes to employ disorder as a knob that spatially varies parameters such as doping and strain within a single sample, to test and quantify the relationship of CO to SC. However, this strategy rests on the premise that what is seen on the surface is not
merely a surface effect, but is reflective of the bulk of the sample.
In \BSCO\, while CO has been observed in the bulk (via, {\em e.g.}, resonant X-ray scattering \cite{CominScience2014}) with the same average wavevector as on the surface, it has not yet been demonstrated that they are locally the same phenomenon. Here, we 
combine a local probe, scanning tunneling microscopy (STM), with  
a theoretical framework known as cluster analysis \cite{PhillabaumNatComm2012}, appropriate near a critical point, in order to test whether the surface CO is connected to the bulk CO.
We find that the charge modulations in \BSCO\ have significant stripe character.  
By computing new critical exponents at free surfaces including that of the pair connectivity correlation function, we moreover show 
that these charge modulations pervade the bulk of the sample, and that their spatial correlations are critical throughout the doping range of superconductivity.

We use STM to study the  cuprate high temperature superconductor \BSCO\ (Bi2201) at the dopings shown in Fig.~\ref{fig:phase-diagram}(a), from  underdoped (low $p$) to  overdoped (high $p$) superconducting samples, as well as an optimally doped sample with superconducting transition temperature $T_c=35$ K, as a function of hole concentration $p$. Fig.~\ref{fig:phase-diagram}(b) shows a topographic image of slightly underdoped Bi2201 with transition temperature $T_c = 32$ K (UD32K), drift-corrected as described in Ref.~\cite{LawlerNature2010}. Lead doping suppresses the structural supermodulation, leaving only the atomic corrugations with a periodicity of $a_0 = 3.8$ \AA\ between copper atoms in the Cu-O planes.

\section*{Results}

\begin{figure*}
\includegraphics[width=1.5\columnwidth]{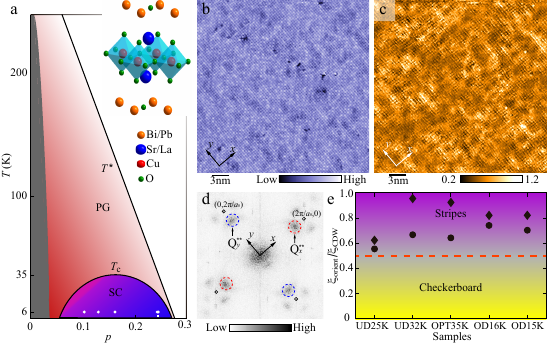}
\caption{\textbf{Phase diagram and stripe order.} (\textbf{a}) Schematic temperature ($T$) versus chemical-doping ($p$) phase diagram of Bi2201, displaying both superconducting (SC) and pseudogap (PG) phases. Inset shows the crystal structure of Bi2201. The five white dots and thin lines denote the five samples studied, namely UD25K ($p=0.101$, $T_c=25$ K), UD32K ($p=0.128$, $T_c=32$ K), OPT35K ($p=0.16$, $T_c=35$ K), OD16K ($p=0.241$, $T_c=16$ K), and OD15K ($p= 0.243$, $T_c=15$ K) (from left to right). (\textbf{b}) Constant-current STM topography of UD32K sample acquired at $I=400$ pA and $V_s$ = -200 mV, over a 30 nm $\times$ 30 nm field of view. The arrows corresponds to the two orthogonal Cu-O bond directions throughout this paper. (\textbf{c}) A typical tunneling asymmetry $R$ map taken at 100 mV (i.e. $R(\vec{r}, 100 \mathrm{mV}) = I(\vec{r},100 \mathrm{mV}) / I(\vec{r},-100 \mathrm{mV})$) in the same field of view shown in (\textbf{b}). The brighter color corresponds to larger $R$. A charge modulation with a period of $\sim4a_0$ is evident in real space. (\textbf{d}) Fourier transform (FT) of $R$ map over the entire FOV in (\textbf{c}), with Bragg vectors $(\pm1, 0)2\pi/a_0$ and $(0, \pm1)2\pi/a_0$ marked by black circles. The wavevectors Q$_{x}^{**}$ $\sim$ (3/4, 0)$2\pi/a_0$ and Q$_{y}^{**}$ $\sim$ (0, 3/4)$2\pi/a_0$ from the charge modulation are identified by dashed circles and arrows. (\textbf{e}) $\xi_{\mathrm{orient}}\mathrm{/}\xi_{\mathrm{CDW}}$ extracted from two different definitions in Refs. \cite{RobertsonPRB2006} (circles) and \cite{DelMaestroPRB2006} (diamonds). All $\xi_{\mathrm{orient}}\mathrm{/}\xi_{\mathrm{CDW}}$ appear greater than 0.5 for various samples, consistent with a striped nature of the charge order. The purple and yellow regions indicate the stripe and checkerboard phases, respectively.
\label{fig:phase-diagram}}
\end{figure*}

\noindent \textbf{Stripes vs.\ Checkers}.
To identify the nature of the charge modulations, we focus on the $R$-map, where $R(\vec{r},V) = I(\vec{r},V)/I(\vec{r},-V)$, and $I(\vec{r},\pm V)$ represents the STM tunneling current at $\pm V$ as a function of position $\vec{r}$ along the surface of the sample \cite{LawlerNature2010, KohsakaScience2007}. The \textit{R}-map has the advantage that it cancels out certain unmeasurable quantities, such as the tunneling matrix element and tunnel barrier height. Fig.~\ref{fig:phase-diagram}(c) shows the $R$-map with $V = 100$ mV in the same field of view (FOV) as Fig.~\ref{fig:phase-diagram}(b). A local modulation with period near $4a_0$ is readily apparent, as confirmed by the two-dimensional Fourier transform (FT) of the $R$-map in Fig.~\ref{fig:phase-diagram}(d), showing peaks at Q$_{x}^{**}$ $\sim$ ($\pm$3/4, 0)$2\pi/a_{0}$ and Q$_{y}^{**}$ $\sim$ (0, $\pm$3/4)$2\pi/a_{0}$.   The Q$^{**}$ peaks carry information about the same charge modulation as the peaks at Q$_{x}^{*}$ $\sim$ ($\pm$1/4, 0)$2\pi/a_{0}$ and Q$_{y}^{*}$ $\sim$ (0, $\pm$1/4)$2\pi/a_{0}$ \cite{ParkerNature2010, FujitaPNAS2014}, and because they are well-separated from the central broad FT peak, there is less measurement error associated with tracking the Q$^{**}$. We therefore focus on the Q$^{**}$ peaks.

There has been experimental evidence in various families of cuprate superconductors for both stripe order (a unidirectional CDW) \cite{TranquadaNature1995, MookNature1998, HowaldPNAS2003,CominScience2015} and checkerboard order (a bidirectional CDW) \cite{HoffmanScience2002vortex,HowaldPRB2003,VershininScience1995,ArpaiaScience2020}. In a real material where quenched disorder is always present, it is difficult to discern from direct observation which tendency (stripes or checkerboards) would dominate in a hypothetical zero disorder limit \cite{RobertsonPRB2006, DelMaestroPRB2006}. The quenched disorder that is always present in real materials can favor the appearance of stripe correlations \cite{DelMaestroPRB2006}. One metric for distinguishing whether the underlying electronic tendency favors stripes or checkerboards is to compare the correlation length of the periodic density modulations $\xi_{\mathrm{CDW}}$ with the correlation length of the orientation of the modulations $\xi_{\mathrm{orient}}$.  Two different theoretical approaches \cite{RobertsonPRB2006, DelMaestroPRB2006} predict that $\xi_{\mathrm{orient}} > \frac{1}{2} \xi_{\mathrm{CDW}}$ when the underlying tendency is toward stripes rather than checkerboard modulations.

In order to infer whether the charge modulations would tend toward stripes or checkerboards in Bi2201 in a hypothetical zero disorder limit, we construct the local Fourier components of the $R$-map at wavevector $\vec{q}$,
\begin{equation}
A(\vec{q},\vec{r})= \frac{1}{2\pi^2 L^2}     \int R(\vec{r})e^{i\vec{q}\cdot \vec{r}'}e^{-(\vec{r} - \vec{r}')^2/2\pi L^2} d^2\vec{r}'
\label{eqn:FT}
\end{equation}
Throughout the paper, we use $L=0.6a_0$ for all $R$-map datasets.
The correlation lengths $\xi_{\rm CDW}$ and $\xi_{\rm orient}$ are then formed from the scalar fields $A_x(\vec{r})=A(Q_x^{**}, \vec{r})$ and $A_y(\vec{r})=A(Q_y^{**}, \vec{r})$ using two different methods as described in Refs.~\cite{RobertsonPRB2006, DelMaestroPRB2006}. Fig.~\ref{fig:phase-diagram}(e) summarizes the ratio of $\xi_{\mathrm{orient}}$/$\xi_{\mathrm{CDW}}$ obtained from each dataset.  In every sample, both methods reveal that Bi2201 tends more towards stripes than checkerboards, since $\xi_{\mathrm{orient}}$/$\xi_{\mathrm{CDW}} > 0.5$.
This reveals that there is significant local stripe order in the system, and that it likely would also be present in the zero disorder limit. Regardless of whether the tendency to stripe modulations survives the hypothetical zero disorder limit, in the material under consideration, the above analyses show that there are local stripe domains present.



\begin{figure*}
\center
 \includegraphics[width=1.5\columnwidth]{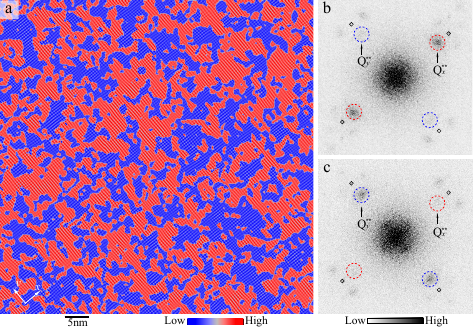}
  \caption{\textbf{Mapping Ising domains.} (\textbf{a}) A 65 nm $\times$ 65 nm Fourier-filtered $R(\vec{r}, 100\,\mathrm{mV})$ map of UD32K, colored red ($\sigma=1$, $Q_{x}^{**}$ dominates) and blue ($\sigma=-1$, $Q_{y}^{**}$ dominates) to indicate the local $Q^{**}$ unidirectional orientations. The unidirectional domains are derived from peaks around $Q_{x}^{**}$ and $Q_{y}^{**}$ with details in the text. The $R$ map has been Fourier-filtered to include only the power spectral density surrounding the four $Q^{**}$ peaks (dashed circles in Fig.\ 1d). (\textbf{b, c}) Fourier transform of red- and blue-colored regions of non-filtered $R$ map in (a), respectively. $Q_{x}^{**}$ dominates in the red regions (\textbf{b}), whereas $Q_{y}^{**}$ in the blue regions (\textbf{c}).
\label{fig:UD32K-mapping}
 }
  \end{figure*}

\noindent \textbf{Ising Domains}.
Having identified the stripe nature of the local charge modulations in Bi2201, we map out where in the sample there are locally $x$-oriented domains, and where there are locally $y$-oriented domains. In Figure~\ref{fig:UD32K-mapping}, we show this mapping for the UD32K sample, constructed as follows: At each position $\vec{r}$, the local FT $A(\vec{q},\vec{r})$ is calculated according to Eqn.~\ref{eqn:FT}, which employs a Gaussian window of width $L$, with $L$ optimized as described in Ref.\ \cite{Supplement}. We then integrate the FT intensity in a 2D gaussian window centered on $\vec{q}=\pm Q_x^{**}$, and divide it by the integrated FT intensity around $\vec{q}=\pm Q_y^{**}$. If this ratio is greater than a threshold $f \sim 1$ ({\em i.e.}\ $Q_x^{**}$ is dominant), the region is colored red in Fig.~\ref{fig:UD32K-mapping}(a), otherwise the region is colored blue. The pattern thus derived in Fig.~\ref{fig:UD32K-mapping}(a) is largely insensitive to changes in detail such as the exact center of the integration window, the size of the integration window, and the threshold $f$ by which a cluster is colored. Similar results are also obtained in other samples with different chemical doping $p$ by quantifying the FT intensity around $Q^{**}$ and $Q^{*}$ (Supplementary Fig.~\ref{fig:Ising-Maps}).

We analyze the pattern formation under the assumption that it is driven by a critical point under the superconducting dome. At the critical point of a second order phase transition, a system exhibits correlated fluctuations on all length scales, resulting in power law behavior for measurable quantities, with a different ``critical exponent'' controlling the power law of each quantity. 

 If the complex pattern formation shown in Fig.~\ref{fig:UD32K-mapping} is due to proximity to a critical point, then the critical exponents would be encoded in the geometric pattern, and the quantitative characteristics of the clusters would act like a  fingerprint to identify the critical point controlling the pattern formation. This reveals information such as the relative importance of disorder and interactions. 
Because critical exponents are particularly sensitive to dimension, this analysis can also reveal whether the
clusters form only on the surface of the material (like frost on a window), or whether they extend seamlessly from the surface into the bulk (like a tree whose roots reach deep underground). Unless the structures seen on the surface pervade the bulk of the material, they
cannot be responsible 
for the  bulk superconductivity. \\

\begin{figure}
  \includegraphics[width=0.95\columnwidth]{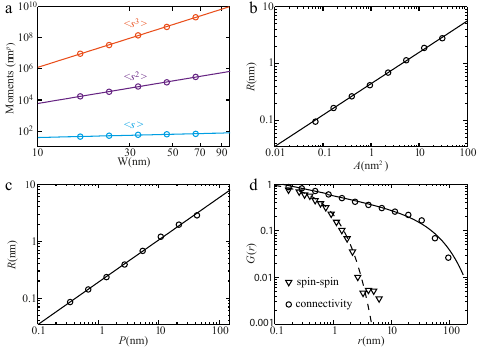}
  \caption{\textbf{Cluster structure and power-law statistical analysis in UD32K sample.} (\textbf{a}) Finite-size-scaling of moments for cluster size distribution, from which the Fisher exponent $\tau$ is calculated. Here p corresponds to the power indexes for the first ($p$ = 1), second ($p$ = 2) and third ($p$ = 3) moments. (\textbf{b}) The radius of gyration \textit{R} versus cluster area \textit{A} showing a power law between them, from which the critical exponent $d_v^*$ is extracted.
  (\textbf{c}) The radius of gyration \textit{R} versus effective cluster perimeter \textit{P}.  The perimeter \textit{P} also shows a power law dependence on \textit{R},  
 from which the critical exponent $d_h^*$ is extracted.
   (\textbf{d}) Spatial correlation functions $G_{\textrm{conn}}(\textbf{r})$ (circles) and $G_{\textrm{spin}}(\textbf{r})$ (triangles) for calculating the critical exponent $d-2+\eta_{||}$. The black line shows the best fit of the pair connectivity correlation function by $G_{\textrm{conn}}\propto r^{-(d-2+\eta_{||})}\textrm{exp}(-r\textrm{/}\xi\textrm{)}$, whereas the dashed line is only a guide to the eye. Logarithmic binning has been used in (\textbf{b-d}).\cite{NewmanContPhys2005}
\label{fig:power-law-fits}}
\end{figure}

\noindent \textbf{Critical Exponents}.
Near a critical point, the number of clusters $D$ of a particular size $s$ is power-law distributed, $D(s) \propto s^{-\tau}$, where $S$ is the number of sites in the cluster and $\tau$ is the Fisher critical exponent \cite{Fisher1967}.
Figure~\ref{fig:power-law-fits}(a)  shows the first $\langle s\rangle$, second $\langle s^2\rangle$, and third $\langle s^3\rangle$ moments of the cluster size distribution as a function of the window size $W$, where $s$ is the observed area of each cluster. Consistent with a system near criticality, the behavior of the moments vs.\ window size $W$ displays robust power law behavior. The cluster moments are related to critical exponents by $\left< s^n \right> \propto W^{(n+1-\tau)d_v^*}$, where $d_v^{*}$ is the effective volume fractal dimension. Since the first moment $\langle s\rangle$ depends only weakly on $W$ (leading to larger error in the estimate of the power law), we combine the information from $\langle s^2\rangle$ and $\langle s^3\rangle$ to derive $\tau$. In the UD32K sample (Fig.~\ref{fig:power-law-fits}(a)), we find $\tau=1.88\pm0.08$.


The boundaries of clusters become fractal in the vicinity of a critical point, scaling as $H \propto R^{d_h}$ where $H$ is the size of each cluster's hull (outer perimeter), $R$ is the radius of gyration of each cluster, and $d_h$ is the fractal dimension of the hull. The interiors of the clusters also become fractal, scaling as $V \propto R^{d_v}$, where $d_v$ is the volume fractal dimension of the clusters. Since STM probes only the sample surface, the observable quantities are the area $A \propto R^{d_v^*}$ and perimeter $P \propto R^{d_h^*}$ of each cluster, where $d_v^*$ and $d_h^*$ represent the effective volume and hull fractal dimension, respectively. 
In Fig.~\ref{fig:power-law-fits} (b) and (c), the cluster properties $A$ and $P$ are plotted vs. $R$, revealing a robust power law spanning 2.5 decades for both $d_h^*$ and $d_v^*$. 
Using a straightforward linear fit of the log-log plots (where the first point is omitted from the fit, since short-distance fluctuations are nonuniversal), we obtain the critical exponents $d_v^*=1.82\pm 0.03$ [Fig.~\ref{fig:power-law-fits}(b)] and $d_h^*=1.35\pm 0.03$ [Fig.~\ref{fig:power-law-fits}(c)].

We turn now to the orientation-orientation correlation function $G_{\mathrm{orient}}(\vec{r})$, which is the analogue of the spin-spin correlation function familiar from Ising models, $G_{\mathrm{orient}}(\vec{r}) = G_{\mathrm{spin}}(\vec{r})=\langle S(\vec{r})S(0)\rangle-\langle S(0)\rangle^2$, where $\vec{r}=|\vec{R}_i-\vec{R}_j|$ is the distance between $(x,y)$ positions, here measured only on the surface. Near criticality, this function displays power law behavior as $G_{\rm orient}(r) \propto 1/r^{d-2+\eta_{||}}$, where $\eta_{||}$ is the anomalous dimension as measured at the surface, and $d$ is the physical dimension of the phenomenon being studied, whether $d=2$ for a surface phenomenon, or $d=3$ for physics arising from the bulk interior of the material.
Figure~\ref{fig:power-law-fits}(d)  shows $G_{\mathrm{orient}}(\vec{r})$ for UD32K (triangles). For the UD32K sample as well as the other  samples studied, $G_{\mathrm{orient}}(\vec{r})$ does not have the standard power-law behavior expected near a critical point, but instead it decays more quickly with $\vec{r}$.

Whereas the orientation-orientation correlation function is not power law in the data, the {\em pair connectivity function}, which is the probability that two
aligned 
regions a distance $r$ apart are in the same connected cluster \cite{StaufferBook1994},
does display robust power law behavior in the data, with $G_{\mathrm{conn}}(r) \propto r^{-(d-2+\eta_{\mathrm{conn}})}$, 
with $d-2+\eta_{\mathrm{conn}} = 0.29 +/- 0.036$,
as shown in Fig.~\ref{fig:power-law-fits}(d) (circles).     
 
While the pair connectivity function has been widely discussed for uncorrelated percolation fixed points \cite{StaufferBook1994}, where it is a power law, it has not previously been characterized at other fixed points. Our simulations of both the clean and random field Ising models show that the pair connectivity function is {\em also} a power law at the 
2D clean Ising (C-2D) and 
the 2D random field Ising (RF-2D) fixed points (see SI). We find that it also displays power law behavior on interior 2D slices\cite{C-3Dx} and at a free surface for the 
3D clean Ising (C-3D) and 3D random field Ising (RF-3D) fixed points. In addition, our simulations  of the clean and random field models close to but not at criticality show that there is a regime in which  a short correlation length $\xi_{\rm spin}$ is evident in the spin-spin correlation function, 
in conjunction with robust power law behavior with a long correlation length $\xi_{\mathrm{cluster}}$  in the pair connectivity function, consistent with this dataset (see SI).  

\begin{figure}
\center
 \includegraphics[width=1\columnwidth]{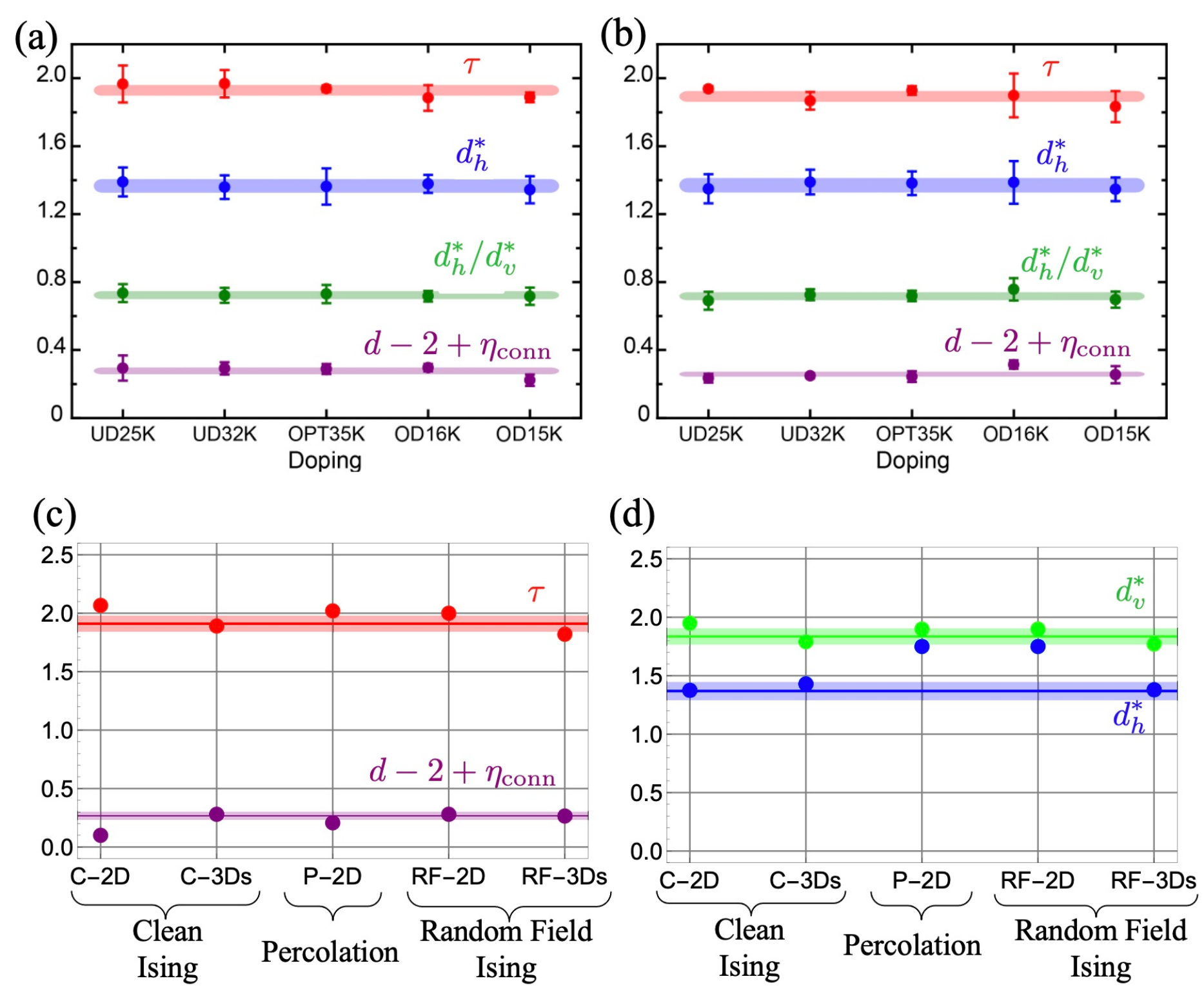}
  \caption{\textbf{Critical exponents.} (a) and (b)   Experimental values of critical exponents derived from Q$^{**}$  and Q$^{*}$ cluster maps, respectively.    For the critical exponent $\tau$, the error bars are estimated as the standard deviation of $\tau$ calculated from different gaussian width $L$ and crop size W.  For the critical exponents  $d_v^*$, $d_h^*\textrm{/}d_v^*$,  and $d-2+\eta_{||}$, the statistical errors are estimated as the standard deviation of the critical exponents calculated from different gaussian width $L$ and logarithmic binning.
No obvious chemical doping dependence is observed, indicative of 
generic critical stripe correlations
in Bi2201.
Thick lines represent the spread of critical exponents for various samples.
(c) Comparison of data-derived exponents to theoretical values of candidate fixed points for exponents $\tau$ and $d-2+\eta_{\rm connect}$.  (d)   Comparison of data-derived exponents to theoretical values of candidate fixed points for exponents $d_v^*$ and $d_h^*$.
In both (c) and (d), circles represent theoretical values, and the thick
line represents the spread of critical exponents over all dopings.
}
\label{fig:exponent-comparison}
\end{figure}

Figure~\ref{fig:exponent-comparison}(c) and (d) show a comparison between the data-derived critical exponents from the R-map of UD32K, and the theoretical critical exponents of Eqn.~\ref{eqn:model}. 
The data-derived value of $d_h^*$ is inconsistent with the 2D percolation (P-2D) fixed point, indicating that interactions between stripe orientations must be present.  In addition, the data-derived value for $d-2+\eta_{\rm conn}$ is inconsistent with that of the C-2D fixed point, and the data-derived value of $d_h^*$ is inconsistent with the RF-2D fixed point.  The remaining candidate fixed points controlling the power law order of stripe orientations are C-3D and RF-3D (denoted C-3Ds and RF-3Ds, respectively, in the figure because we report theoretical values of the exponents at a free surface of the 3D models).  Therefore, we find that the data-derived exponents are consistent with those of a layered clean or random field Ising model with  $J^{\perp} > 0$ near criticality. 

This shows that the fractal patterns observed here via STM are not confined to the surface, like frost growing on a window pane.  Rather, these fractal stripe clusters fill the bulk of the material, more like a tree viewed through a 2D window.   
In the same way that transverse stripe fluctuations 
help electron pairs condense into a superconducting state rather
than into the competing (insulating) pair crystal phase,\cite{concepts,EmeryPRB1997,KivelsonNature1998}
the stripe orientation fluctuations observed here 
could also have a profound effect on superconductivity, 
since orientation fluctuations of stripes also frustrate the pair crystal.

We find similar results on the samples at other dopings, which also show robust power laws, with the same exponents within error bars, and a cluster correlation length $\xi_{\mathrm{cluster}}$ which exceeds the FOV (see SI). The doping independence is surprising, since one would expect there to be a phase transition from ordered to disordered stripe orientations as doping (a source of quenched disorder) is increased, with the critical, power law behavior observed here limited to the vicinity of the phase transition.  While a broad region of critical behavior like that observed here is not natural near the C-3D fixed point, a broad region of critical behavior is {\em characteristic} of the RF-3D fixed point: 
for example, the cluster size distribution $D(S)$ displays 2 decades of scaling, 50\% away from the RF-3D critical point\cite{PerkovicPRL1995}. 

\section*{Discussion}
While our findings suggest a prominent role for criticality in the phase diagram of cuprate superconductors, the spatial structures reported here are inconsistent with {\em quantum} criticality because these correlations are static on the timescales of several seconds,
whereas quantum critical correlations fluctuate in time.
In addition, for a quantum critical point tuned by doping, quantum critical scaling is confined to a narrow region close to the critical doping, in a 
``wedge'' emanating from the critical doping and extending up in temperature.
By contrast, we find critical, power law correlations at low temperature {\em throughout the entire doping range measured}.  Whereas the lack of detectable doping dependence in finite FOV's is inconsistent with quantum criticality, it is natural in the classical, three-dimensional random field Ising model. 
What the RF-3D fixed point shares in common with quantum criticality is that it is also a zero temperature critical point. However, it is tuned by disorder rather than by quantum fluctuations.

In addition, RF-3D is notoriously difficult to equilibrate in the vicinity of the critical point, since the relaxation time scales exponentially with the spin-spin correlation length: $\tau_{\rm rel} \sim \exp[\xi_{\mathrm{spin}}^\theta]$, where the violation of hyperscaling exponent $\theta = 1.5$~.  
If the spin-spin correlation length reaches even 10 unit cells, the relaxation time will be ${\rm exp}[10^{1.5}]\approx 5\times 10^{13}$ times any bare microscopic timescale.  Compare this with critical scaling near the C-3D fixed point, where for a correlation length of 10 unit cells, the relaxation time is on the order of $\tau_{\rm rel} \propto \xi_{\rm spin}^{z} \approx 10$ times the bare timescale.\footnote{Here, $z$ is the dynamical critical exponent, which is of order 1.}  

As temperature is lowered on any given sample, it falls out of equilibrium if the relaxation time exceeds a timescale $t_0$ which is set by the cooling protocol, 
$\tau_{rel} \propto {\rm exp}[\xi_{\rm spin}^\theta] \gtrsim t_0$.
Thus the orientational correlation length depends on the cooling protocol, rather than on doping, when approaching an ordered ground state ({\em i.e.} for doping $p < p_c$), where the correlation length scales as $\xi_{\rm spin} \propto 1/|T-T_c|^{\nu}$.  Beyond that doping, the correlation length scales as $\xi_{\rm spin} \propto 1/|p-p_c|^{\nu}$ at low temperature.  
Experiments at larger FOV and higher doping could determine whether there is a doping dependence to the cluster correlation length $\xi_{\rm cluster}$, and  thereby identify the critical doping concentration for the vestigial nematic \cite{NiePNAS2014}.

Our discovery that the charge modulations observed at the surface are locally one dimensional and also extend throughout the bulk of the material has important implications for the mechanism of superconductivity in these materials.
The fractal stripe clusters may have a profound effect on superconductivity, by frustrating competing orders like the pair crystal.
The cluster analysis framework demonstrated here extends the capability of all surface probes used to study quantum materials to distinguish surface from bulk behavior.   Furthermore, our finding that fractal stripe patterns both permeate the bulk of a cuprate superconductor and that they share universal features throughout the superconducting dome, raises  important questions.  Because doping naturally introduces disorder, a disorder-driven, zero temperature critical point for electronic nematicity is a very real possibility in other cuprates as well.  More work is needed to further elucidate the connection between these fractal electronic textures and superconductivity.  For example, the connectivity correlation length of the stripes exceeds the field of view of our experiments throughout the doping range.  An important open question for future studies is to establish the relationship between this correlation length and the optimal superconducting transition temperature.    \\

\section*{Methods}
\noindent \textbf{STM measurements}.
Two different home-built STMs were used to acquire the data in this paper, both in cryogenic ultra-high vacuum. The samples were cleaved \textit{in situ} at $\sim$25 K and inserted immediately into the STM sample stage for imaging at 6 K. A mechanically cut polycrystalline PtIr tip was firstly calibrated in Au single crystals to eliminate large tip anisotropy. To obtain a tunneling current, we applied a bias to the sample while the tip was held at virtual ground. All tunneling spectra, which are proportional to the local density of states at given sample voltage, were measured using a standard lock-in technique. \\

\noindent \textbf{Theoretical Models}.
Because there are only two orientations of the unidirectional domains, we can map the orientations to an Ising variable \cite{PhillabaumNatComm2012, CarlsonPRL2006, CarlsonJSNM2015}, $\sigma = \pm 1$, where the $+$ sign corresponds to red regions in Fig.~\ref{fig:UD32K-mapping}, and the $-$ sign corresponds to the blue regions. We model the tendency of neighboring unidirectional regions to align by a ferromagnetic interaction within each plane $J^{||}$ as well as an interlayer coupling $J^{\perp}$
\begin{equation}
H = - \sum_{\left< i j \right>_{||}} J^{||}\sigma_i \sigma_j - \sum_{\left< i j \right>_{\perp}}J^{\perp} \sigma_i \sigma_j - \sum_i (h_i+h) \sigma_i~.
\label{eqn:model}
\end{equation}
Any net orienting field, whether applied or intrinsic to the crystal, contributes to the bulk orienting field $h$ \cite{PhillabaumNatComm2012}.
In any given region, the local pattern of quenched disorder breaks the rotational symmetry of the host crystal, corresponding to random field disorder $h_i$ in the Ising model \footnote{Quenched disorder can also introduce randomness in the couplings $J$, also known as random bond disorder. In the presence of both random bond and random field disorder, the critical behavior is controlled by the random field fixed point.}.
In the model, $h_i$ is chosen from a gaussian distribution of width $\Delta$,
which quantifies the disorder strength. \\

\noindent \textbf{Simulation Methods}.
When comparing to a 2D model, the effective fractal dimensions  observed at the surface can be compared directly with those of the model, $d_v^*=d_v$, $d_h^*=d_h$. When comparing to a 3D model, we have calculated the cluster critical exponents of the model at a free surface, denoted C-3Ds and RF-3Ds in Fig.~\ref{fig:exponent-comparison}.  

For the clean Ising model in three dimensions, because the fixed point (C-3D) controlling the continuous phase transition is at finite temperature, we use Monte Carlo simulations to generate stripe orientation configurations.  
To calculate the critical exponents of the 3D clean Ising model at a free surface, a 840x840x840 3D clean Ising model with periodic boundary conditions in the x and y direction and open boundary conditions in the z direction was simulated at $T_c = 4.51J$ with 20000 steps of the parallel Metropolis algorithm.  To compare with the finite field of view of the experiments, we average over nine windows of size 256x256, taken from a free surface of the final spin configuration. 
The averages of these critical exponents are shown in Figure~\ref{fig:exponent-comparison}.  The standard deviations are smaller than the symbol size. 

For the random field Ising model in three dimensions, the fixed point (RF-3D) controlling the continuous phase transition is at zero temperature, and we use a mapping to the max-flow min-cut algorithm \cite{1056816,GoldbergAndrewVTPAf,PicardJ.C1975Mcar} to calculate exact ground state spin orientation configurations.  
The critical point of the 3D random field Ising model occurs at zero temperature. To calculate the 3D random field Ising model surface exponents, ground states were computed for 10 different disorder configurations of a 512x512x512 3D RFIM with open boundary conditions in the z direction and periodic boundary conditions in the x and y directions with $R=3$, using a mapping between RFIM and the max-flow/min-cut problem.\cite{1056816,GoldbergAndrewVTPAf,PicardJ.C1975Mcar} To compare with the finite field of view of the experiments, the top surface of these ground states was windowed to system size 256x256 and critical exponents were extracted from the corresponding windows. 
The averages of these critical exponents are shown in Figure~\ref{fig:exponent-comparison}.  The standard deviations are smaller than the symbol size.  \\

\textbf{Cluster Methods}.
While the exponent $\tau$ derived from STM data is close to the narrow range allowed by the theoretical models, it is slightly below this range.  Estimates of this exponent derived from a finite FOV are known to be skewed toward low values due to a bump in the scaling function, especially in the presence of random field effects\cite{PerkovicPRL1995}. To mitigate this effect, we perform finite-size scaling by analyzing the data as a function of window size $W$.

To mitigate possible window effects associated with a finite FOV in deriving the fractal dimensions d$_h$ and d$_v$, only the internal clusters that touch no edge of the Ising map have been included in the analyses of experimental data as well as simulation results. To extract the effective fractal dimensions, we adopt a standard logarithmic binning technique for analyzing  power-law behavior \cite{NewmanContPhys2005}.

\begin{acknowledgements}
We thank Michael Boyer for sharing part of the data discussed herein.  The authors thank S.A.\ Kivelson for helpful discussions.  
C.L.S.\ acknowledges support by the Golub Fellowship at Harvard University. 
S.L., B.P., and E.W.C. acknowledge support from National Science Foundation Grant No. DMR-1508236 and Department of Education Grant No. P116F140459.  S.L.  acknowledges support from a Bilsland Dissertation Fellowship. 
F.S. and E.W.C. acknowledge support from NSF Grant No. DMR-2006192,  a Research Corporation for Science Advancement SEED Award, and XSEDE Grant Nos. TG-DMR-180098 and DMR-190014.  E.W.C. acknowledges support from a Fulbright Fellowship, and thanks the Laboratoire de Physique et d'\'{E}tude des Mat\'{e}riaux (LPEM) at \'{E}cole Sup\'{e}rieure de Physique et de Chimie Industrielles de la Ville de Paris (ESPCI) for hospitality. K.A.D. acknowledges support from NSF Grant No. CBET-1336634. This research was supported in part through computational resources provided by Information Technology at Purdue, West Lafayette, IN.\cite{rcac-purdue}
\end{acknowledgements}

\bibliography{criticalexponents}

\providecommand{\noopsort}[1]{}\providecommand{\singleletter}[1]{#1}%
\begin{thebibliography}{56}%
\makeatletter
\providecommand \@ifxundefined [1]{%
 \@ifx{#1\undefined}
}%
\providecommand \@ifnum [1]{%
 \ifnum #1\expandafter \@firstoftwo
 \else \expandafter \@secondoftwo
 \fi
}%
\providecommand \@ifx [1]{%
 \ifx #1\expandafter \@firstoftwo
 \else \expandafter \@secondoftwo
 \fi
}%
\providecommand \natexlab [1]{#1}%
\providecommand \enquote  [1]{``#1''}%
\providecommand \bibnamefont  [1]{#1}%
\providecommand \bibfnamefont [1]{#1}%
\providecommand \citenamefont [1]{#1}%
\providecommand \href@noop [0]{\@secondoftwo}%
\providecommand \href [0]{\begingroup \@sanitize@url \@href}%
\providecommand \@href[1]{\@@startlink{#1}\@@href}%
\providecommand \@@href[1]{\endgroup#1\@@endlink}%
\providecommand \@sanitize@url [0]{\catcode `\\12\catcode `\$12\catcode
  `\&12\catcode `\#12\catcode `\^12\catcode `\_12\catcode `\%12\relax}%
\providecommand \@@startlink[1]{}%
\providecommand \@@endlink[0]{}%
\providecommand \url  [0]{\begingroup\@sanitize@url \@url }%
\providecommand \@url [1]{\endgroup\@href {#1}{\urlprefix }}%
\providecommand \urlprefix  [0]{URL }%
\providecommand \Eprint [0]{\href }%
\providecommand \doibase [0]{https://doi.org/}%
\providecommand \selectlanguage [0]{\@gobble}%
\providecommand \bibinfo  [0]{\@secondoftwo}%
\providecommand \bibfield  [0]{\@secondoftwo}%
\providecommand \translation [1]{[#1]}%
\providecommand \BibitemOpen [0]{}%
\providecommand \bibitemStop [0]{}%
\providecommand \bibitemNoStop [0]{.\EOS\space}%
\providecommand \EOS [0]{\spacefactor3000\relax}%
\providecommand \BibitemShut  [1]{\csname bibitem#1\endcsname}%
\let\auto@bib@innerbib\@empty
\bibitem [{\citenamefont {Hoffman}\ \emph {et~al.}(2002)\citenamefont
  {Hoffman}, \citenamefont {Hudson}, \citenamefont {Lang}, \citenamefont
  {Madhavan}, \citenamefont {Eisaki}, \citenamefont {Uchida},\ and\
  \citenamefont {Davis}}]{HoffmanScience2002vortex}%
  \BibitemOpen
  \bibfield  {author} {\bibinfo {author} {\bibfnamefont {J.~E.}\ \bibnamefont
  {Hoffman}}, \bibinfo {author} {\bibfnamefont {E.~W.}\ \bibnamefont {Hudson}},
  \bibinfo {author} {\bibfnamefont {K.~M.}\ \bibnamefont {Lang}}, \bibinfo
  {author} {\bibfnamefont {V.}~\bibnamefont {Madhavan}}, \bibinfo {author}
  {\bibfnamefont {H.}~\bibnamefont {Eisaki}}, \bibinfo {author} {\bibfnamefont
  {S.}~\bibnamefont {Uchida}},\ and\ \bibinfo {author} {\bibfnamefont {J.~C.}\
  \bibnamefont {Davis}},\ }\bibfield  {title} {\bibinfo {title} {A four unit
  cell periodic pattern of quasi-particle states surrounding vortex cores in
  {Bi$_2$Sr$_2$CaCu$_2$O$_{8+d}$}},\ }\href
  {https://doi.org/10.1126/science.1066974} {\bibfield  {journal} {\bibinfo
  {journal} {Science}\ }\textbf {\bibinfo {volume} {295}},\ \bibinfo {pages}
  {466} (\bibinfo {year} {2002})}\BibitemShut {NoStop}%
\bibitem [{\citenamefont {Howald}\ \emph
  {et~al.}(2003{\natexlab{a}})\citenamefont {Howald}, \citenamefont {Eisaki},
  \citenamefont {Kaneko},\ and\ \citenamefont {Kapitulnik}}]{HowaldPNAS2003}%
  \BibitemOpen
  \bibfield  {author} {\bibinfo {author} {\bibfnamefont {C.}~\bibnamefont
  {Howald}}, \bibinfo {author} {\bibfnamefont {H.}~\bibnamefont {Eisaki}},
  \bibinfo {author} {\bibfnamefont {N.}~\bibnamefont {Kaneko}},\ and\ \bibinfo
  {author} {\bibfnamefont {A.}~\bibnamefont {Kapitulnik}},\ }\bibfield  {title}
  {\bibinfo {title} {Coexistence of periodic modulation of quasiparticle states
  and superconductivity in {Bi$_2$Sr$_2$CaCu$_2$O$_{8+d}$}},\ }\href
  {https://doi.org/10.1073/pnas.1233768100} {\bibfield  {journal} {\bibinfo
  {journal} {Proceedings of the National Academy of Sciences}\ }\textbf
  {\bibinfo {volume} {100}},\ \bibinfo {pages} {9705} (\bibinfo {year}
  {2003}{\natexlab{a}})}\BibitemShut {NoStop}%
\bibitem [{\citenamefont {Kohsaka}\ \emph {et~al.}(2007)\citenamefont
  {Kohsaka}, \citenamefont {Taylor}, \citenamefont {Fujita}, \citenamefont
  {Schmidt}, \citenamefont {Lupien}, \citenamefont {Hanaguri}, \citenamefont
  {Azuma}, \citenamefont {Takano}, \citenamefont {Eisaki}, \citenamefont
  {Takagi}, \citenamefont {Uchida},\ and\ \citenamefont
  {Davis}}]{KohsakaScience2007}%
  \BibitemOpen
  \bibfield  {author} {\bibinfo {author} {\bibfnamefont {Y.}~\bibnamefont
  {Kohsaka}}, \bibinfo {author} {\bibfnamefont {C.}~\bibnamefont {Taylor}},
  \bibinfo {author} {\bibfnamefont {K.}~\bibnamefont {Fujita}}, \bibinfo
  {author} {\bibfnamefont {A.~R.}\ \bibnamefont {Schmidt}}, \bibinfo {author}
  {\bibfnamefont {C.}~\bibnamefont {Lupien}}, \bibinfo {author} {\bibfnamefont
  {T.}~\bibnamefont {Hanaguri}}, \bibinfo {author} {\bibfnamefont
  {M.}~\bibnamefont {Azuma}}, \bibinfo {author} {\bibfnamefont
  {M.}~\bibnamefont {Takano}}, \bibinfo {author} {\bibfnamefont
  {H.}~\bibnamefont {Eisaki}}, \bibinfo {author} {\bibfnamefont
  {H.}~\bibnamefont {Takagi}}, \bibinfo {author} {\bibfnamefont
  {S.}~\bibnamefont {Uchida}},\ and\ \bibinfo {author} {\bibfnamefont {J.~C.}\
  \bibnamefont {Davis}},\ }\bibfield  {title} {\bibinfo {title} {An intrinsic
  bond-centered electronic glass with unidirectional domains in underdoped
  cuprates},\ }\href {https://doi.org/10.1126/science.1138584} {\bibfield
  {journal} {\bibinfo  {journal} {Science}\ }\textbf {\bibinfo {volume}
  {315}},\ \bibinfo {pages} {1380} (\bibinfo {year} {2007})}\BibitemShut
  {NoStop}%
\bibitem [{\citenamefont {Wise}\ \emph {et~al.}(2008)\citenamefont {Wise},
  \citenamefont {Boyer}, \citenamefont {Chatterjee}, \citenamefont {Kondo},
  \citenamefont {Takeuchi}, \citenamefont {Ikuta}, \citenamefont {Wang},\ and\
  \citenamefont {Hudson}}]{WiseNatPhys2008}%
  \BibitemOpen
  \bibfield  {author} {\bibinfo {author} {\bibfnamefont {W.~D.}\ \bibnamefont
  {Wise}}, \bibinfo {author} {\bibfnamefont {M.~C.}\ \bibnamefont {Boyer}},
  \bibinfo {author} {\bibfnamefont {K.}~\bibnamefont {Chatterjee}}, \bibinfo
  {author} {\bibfnamefont {T.}~\bibnamefont {Kondo}}, \bibinfo {author}
  {\bibfnamefont {T.}~\bibnamefont {Takeuchi}}, \bibinfo {author}
  {\bibfnamefont {H.}~\bibnamefont {Ikuta}}, \bibinfo {author} {\bibfnamefont
  {Y.}~\bibnamefont {Wang}},\ and\ \bibinfo {author} {\bibfnamefont {E.~W.}\
  \bibnamefont {Hudson}},\ }\bibfield  {title} {\bibinfo {title}
  {Charge-density-wave origin of cuprate checkerboard visualized by scanning
  tunnelling microscopy},\ }\href {https://doi.org/10.1038/nphys1021}
  {\bibfield  {journal} {\bibinfo  {journal} {Nature Physics}\ }\textbf
  {\bibinfo {volume} {4}},\ \bibinfo {pages} {696} (\bibinfo {year}
  {2008})}\BibitemShut {NoStop}%
\bibitem [{\citenamefont {Hanaguri}\ \emph {et~al.}(2004)\citenamefont
  {Hanaguri}, \citenamefont {Lupien}, \citenamefont {Kohsaka}, \citenamefont
  {Lee}, \citenamefont {Azuma}, \citenamefont {Takano}, \citenamefont
  {Takagi},\ and\ \citenamefont {Davis}}]{HanaguriNature2004}%
  \BibitemOpen
  \bibfield  {author} {\bibinfo {author} {\bibfnamefont {T.}~\bibnamefont
  {Hanaguri}}, \bibinfo {author} {\bibfnamefont {C.}~\bibnamefont {Lupien}},
  \bibinfo {author} {\bibfnamefont {Y.}~\bibnamefont {Kohsaka}}, \bibinfo
  {author} {\bibfnamefont {D.~H.}\ \bibnamefont {Lee}}, \bibinfo {author}
  {\bibfnamefont {M.}~\bibnamefont {Azuma}}, \bibinfo {author} {\bibfnamefont
  {M.}~\bibnamefont {Takano}}, \bibinfo {author} {\bibfnamefont
  {H.}~\bibnamefont {Takagi}},\ and\ \bibinfo {author} {\bibfnamefont {J.~C.}\
  \bibnamefont {Davis}},\ }\bibfield  {title} {\bibinfo {title} {A checkerboard
  electronic crystal state in lightly hole-doped
  {Ca$_{2-x}$Na$_x$CuO$_2$Cl$_2$}},\ }\href
  {https://doi.org/10.1038/nature02861} {\bibfield  {journal} {\bibinfo
  {journal} {Nature}\ }\textbf {\bibinfo {volume} {430}},\ \bibinfo {pages}
  {1001} (\bibinfo {year} {2004})}\BibitemShut {NoStop}%
\bibitem [{\citenamefont {Wu}\ \emph {et~al.}(2011)\citenamefont {Wu},
  \citenamefont {Mayaffre}, \citenamefont {Kr{\"{a}}mer}, \citenamefont
  {Horvati{\'{c}}}, \citenamefont {Berthier}, \citenamefont {Hardy},
  \citenamefont {Liang}, \citenamefont {Bonn},\ and\ \citenamefont
  {Julien}}]{WuNature2011}%
  \BibitemOpen
  \bibfield  {author} {\bibinfo {author} {\bibfnamefont {T.}~\bibnamefont
  {Wu}}, \bibinfo {author} {\bibfnamefont {H.}~\bibnamefont {Mayaffre}},
  \bibinfo {author} {\bibfnamefont {S.}~\bibnamefont {Kr{\"{a}}mer}}, \bibinfo
  {author} {\bibfnamefont {M.}~\bibnamefont {Horvati{\'{c}}}}, \bibinfo
  {author} {\bibfnamefont {C.}~\bibnamefont {Berthier}}, \bibinfo {author}
  {\bibfnamefont {W.~N.}\ \bibnamefont {Hardy}}, \bibinfo {author}
  {\bibfnamefont {R.}~\bibnamefont {Liang}}, \bibinfo {author} {\bibfnamefont
  {D.~A.}\ \bibnamefont {Bonn}},\ and\ \bibinfo {author} {\bibfnamefont
  {M.-H.}\ \bibnamefont {Julien}},\ }\bibfield  {title} {\bibinfo {title}
  {Magnetic-field-induced charge-stripe order in the high-temperature
  superconductor {YBa$_2$Cu$_3$O$_y$}},\ }\href
  {https://doi.org/10.1038/nature10345} {\bibfield  {journal} {\bibinfo
  {journal} {Nature}\ }\textbf {\bibinfo {volume} {477}},\ \bibinfo {pages}
  {191} (\bibinfo {year} {2011})}\BibitemShut {NoStop}%
\bibitem [{\citenamefont {Chang}\ \emph
  {et~al.}(2012{\natexlab{a}})\citenamefont {Chang}, \citenamefont {Blackburn},
  \citenamefont {Holmes}, \citenamefont {Christensen}, \citenamefont {Larsen},
  \citenamefont {Mesot}, \citenamefont {Liang}, \citenamefont {Bonn},
  \citenamefont {Hardy}, \citenamefont {Watenphul}, \citenamefont {von
  Zimmermann}, \citenamefont {Forgan},\ and\ \citenamefont
  {Hayden}}]{ChangNatPhys2012}%
  \BibitemOpen
  \bibfield  {author} {\bibinfo {author} {\bibfnamefont {J.}~\bibnamefont
  {Chang}}, \bibinfo {author} {\bibfnamefont {E.}~\bibnamefont {Blackburn}},
  \bibinfo {author} {\bibfnamefont {A.~T.}\ \bibnamefont {Holmes}}, \bibinfo
  {author} {\bibfnamefont {N.~B.}\ \bibnamefont {Christensen}}, \bibinfo
  {author} {\bibfnamefont {J.}~\bibnamefont {Larsen}}, \bibinfo {author}
  {\bibfnamefont {J.}~\bibnamefont {Mesot}}, \bibinfo {author} {\bibfnamefont
  {R.}~\bibnamefont {Liang}}, \bibinfo {author} {\bibfnamefont {D.~A.}\
  \bibnamefont {Bonn}}, \bibinfo {author} {\bibfnamefont {W.~N.}\ \bibnamefont
  {Hardy}}, \bibinfo {author} {\bibfnamefont {A.}~\bibnamefont {Watenphul}},
  \bibinfo {author} {\bibfnamefont {M.}~\bibnamefont {von Zimmermann}},
  \bibinfo {author} {\bibfnamefont {E.~M.}\ \bibnamefont {Forgan}},\ and\
  \bibinfo {author} {\bibfnamefont {S.~M.}\ \bibnamefont {Hayden}},\ }\bibfield
   {title} {\bibinfo {title} {Direct observation of competition between
  superconductivity and charge density wave order in
  {YBa$_2$Cu$_3$O$_{6.67}$}},\ }\href {https://doi.org/10.1038/nphys2456}
  {\bibfield  {journal} {\bibinfo  {journal} {Nature Physics}\ }\textbf
  {\bibinfo {volume} {8}},\ \bibinfo {pages} {871} (\bibinfo {year}
  {2012}{\natexlab{a}})}\BibitemShut {NoStop}%
\bibitem [{\citenamefont {Ghiringhelli}\ \emph
  {et~al.}(2012{\natexlab{a}})\citenamefont {Ghiringhelli}, \citenamefont
  {Le~Tacon}, \citenamefont {Minola}, \citenamefont {Blanco-Canosa},
  \citenamefont {Mazzoli}, \citenamefont {Brookes}, \citenamefont {De~Luca},
  \citenamefont {Frano}, \citenamefont {Hawthorn}, \citenamefont {He},
  \citenamefont {Loew}, \citenamefont {Sawatzky}, \citenamefont {Keimer},\ and\
  \citenamefont {Braicovich}}]{GhiringhelliScience2012}%
  \BibitemOpen
  \bibfield  {author} {\bibinfo {author} {\bibfnamefont {G.}~\bibnamefont
  {Ghiringhelli}}, \bibinfo {author} {\bibfnamefont {M.}~\bibnamefont
  {Le~Tacon}}, \bibinfo {author} {\bibfnamefont {M.}~\bibnamefont {Minola}},
  \bibinfo {author} {\bibfnamefont {S.}~\bibnamefont {Blanco-Canosa}}, \bibinfo
  {author} {\bibfnamefont {C.}~\bibnamefont {Mazzoli}}, \bibinfo {author}
  {\bibfnamefont {N.~B.}\ \bibnamefont {Brookes}}, \bibinfo {author}
  {\bibfnamefont {G.~M.}\ \bibnamefont {De~Luca}}, \bibinfo {author}
  {\bibfnamefont {a.}~\bibnamefont {Frano}}, \bibinfo {author} {\bibfnamefont
  {D.~G.}\ \bibnamefont {Hawthorn}}, \bibinfo {author} {\bibfnamefont
  {F.}~\bibnamefont {He}}, \bibinfo {author} {\bibfnamefont {T.}~\bibnamefont
  {Loew}}, \bibinfo {author} {\bibfnamefont {G.~A.}\ \bibnamefont {Sawatzky}},
  \bibinfo {author} {\bibfnamefont {B.}~\bibnamefont {Keimer}},\ and\ \bibinfo
  {author} {\bibfnamefont {L.}~\bibnamefont {Braicovich}},\ }\bibfield  {title}
  {\bibinfo {title} {Long-range incommensurate charge fluctuations in
  {(Y,Nd)Ba$_2$Cu$_3$O$_{6+x}$}},\ }\href
  {https://doi.org/10.1126/science.1223532} {\bibfield  {journal} {\bibinfo
  {journal} {Science}\ }\textbf {\bibinfo {volume} {337}},\ \bibinfo {pages}
  {821} (\bibinfo {year} {2012}{\natexlab{a}})}\BibitemShut {NoStop}%
\bibitem [{\citenamefont {Comin}\ \emph {et~al.}(2014)\citenamefont {Comin},
  \citenamefont {Eisaki}, \citenamefont {Fra{\~n}o}, \citenamefont {Yee},
  \citenamefont {Yoshida}, \citenamefont {Schierle}, \citenamefont {Weschke},
  \citenamefont {Sutarto}, \citenamefont {He}, \citenamefont {Soumyanarayanan},
  \citenamefont {He}, \citenamefont {Le~Tacon}, \citenamefont {Elfimov},
  \citenamefont {Hoffman}, \citenamefont {Sawatzky}, \citenamefont {Keimer},\
  and\ \citenamefont {Damascelli}}]{CominScience2014}%
  \BibitemOpen
  \bibfield  {author} {\bibinfo {author} {\bibfnamefont {R.}~\bibnamefont
  {Comin}}, \bibinfo {author} {\bibfnamefont {H.}~\bibnamefont {Eisaki}},
  \bibinfo {author} {\bibfnamefont {A.}~\bibnamefont {Fra{\~n}o}}, \bibinfo
  {author} {\bibfnamefont {M.~M.}\ \bibnamefont {Yee}}, \bibinfo {author}
  {\bibfnamefont {Y.}~\bibnamefont {Yoshida}}, \bibinfo {author} {\bibfnamefont
  {E.}~\bibnamefont {Schierle}}, \bibinfo {author} {\bibfnamefont
  {E.}~\bibnamefont {Weschke}}, \bibinfo {author} {\bibfnamefont
  {R.}~\bibnamefont {Sutarto}}, \bibinfo {author} {\bibfnamefont
  {F.}~\bibnamefont {He}}, \bibinfo {author} {\bibfnamefont {A.}~\bibnamefont
  {Soumyanarayanan}}, \bibinfo {author} {\bibfnamefont {Y.}~\bibnamefont {He}},
  \bibinfo {author} {\bibfnamefont {M.}~\bibnamefont {Le~Tacon}}, \bibinfo
  {author} {\bibfnamefont {I.~S.}\ \bibnamefont {Elfimov}}, \bibinfo {author}
  {\bibfnamefont {J.~E.}\ \bibnamefont {Hoffman}}, \bibinfo {author}
  {\bibfnamefont {G.~A.}\ \bibnamefont {Sawatzky}}, \bibinfo {author}
  {\bibfnamefont {B.}~\bibnamefont {Keimer}},\ and\ \bibinfo {author}
  {\bibfnamefont {A.}~\bibnamefont {Damascelli}},\ }\bibfield  {title}
  {\bibinfo {title} {Charge order driven by fermi-arc instability in
  {Bi$_2$Sr$_{2-x}$La$_x$CuO$_{6+\delta}$}},\ }\href
  {https://doi.org/10.1126/science.1242996} {\bibfield  {journal} {\bibinfo
  {journal} {Science}\ }\textbf {\bibinfo {volume} {343}},\ \bibinfo {pages}
  {390} (\bibinfo {year} {2014})}\BibitemShut {NoStop}%
\bibitem [{\citenamefont {da~Silva~Neto}\ \emph {et~al.}(2015)\citenamefont
  {da~Silva~Neto}, \citenamefont {Comin}, \citenamefont {He}, \citenamefont
  {Sutarto}, \citenamefont {Jiang}, \citenamefont {Greene}, \citenamefont
  {Sawatzky},\ and\ \citenamefont {Damascelli}}]{daSilvaNetoScience2015}%
  \BibitemOpen
  \bibfield  {author} {\bibinfo {author} {\bibfnamefont {E.~H.}\ \bibnamefont
  {da~Silva~Neto}}, \bibinfo {author} {\bibfnamefont {R.}~\bibnamefont
  {Comin}}, \bibinfo {author} {\bibfnamefont {F.}~\bibnamefont {He}}, \bibinfo
  {author} {\bibfnamefont {R.}~\bibnamefont {Sutarto}}, \bibinfo {author}
  {\bibfnamefont {Y.}~\bibnamefont {Jiang}}, \bibinfo {author} {\bibfnamefont
  {R.~L.}\ \bibnamefont {Greene}}, \bibinfo {author} {\bibfnamefont {G.~A.}\
  \bibnamefont {Sawatzky}},\ and\ \bibinfo {author} {\bibfnamefont
  {A.}~\bibnamefont {Damascelli}},\ }\bibfield  {title} {\bibinfo {title}
  {Charge ordering in the electron-doped superconductor
  {Nd$_{2-x}$Ce$_x$CuO$_4$}},\ }\href {https://doi.org/10.1126/science.1256441}
  {\bibfield  {journal} {\bibinfo  {journal} {Science}\ }\textbf {\bibinfo
  {volume} {347}},\ \bibinfo {pages} {282} (\bibinfo {year}
  {2015})}\BibitemShut {NoStop}%
\bibitem [{\citenamefont {Croft}\ \emph {et~al.}(2014)\citenamefont {Croft},
  \citenamefont {Lester}, \citenamefont {Senn}, \citenamefont {Bombardi},\ and\
  \citenamefont {Hayden}}]{croft_charge_2014}%
  \BibitemOpen
  \bibfield  {author} {\bibinfo {author} {\bibfnamefont {T.~P.}\ \bibnamefont
  {Croft}}, \bibinfo {author} {\bibfnamefont {C.}~\bibnamefont {Lester}},
  \bibinfo {author} {\bibfnamefont {M.~S.}\ \bibnamefont {Senn}}, \bibinfo
  {author} {\bibfnamefont {A.}~\bibnamefont {Bombardi}},\ and\ \bibinfo
  {author} {\bibfnamefont {S.~M.}\ \bibnamefont {Hayden}},\ }\bibfield  {title}
  {{\selectlanguage {english}\bibinfo {title} {Charge density wave fluctuations
  in la$_{2-x}$sr$_x$cuo$_4$ and their competition with superconductivity}},\
  }\href@noop {} {\bibfield  {journal} {\bibinfo  {journal} {Physical Review
  B}\ }\textbf {\bibinfo {volume} {89}},\ \bibinfo {pages} {224513} (\bibinfo
  {year} {2014})}\BibitemShut {NoStop}%
\bibitem [{\citenamefont {Tabis}\ \emph {et~al.}(2014)\citenamefont {Tabis},
  \citenamefont {Li}, \citenamefont {Tacon}, \citenamefont {Braicovich},
  \citenamefont {Kreyssig}, \citenamefont {Minola}, \citenamefont {Dellea},
  \citenamefont {Weschke}, \citenamefont {Veit}, \citenamefont {Ramazanoglu},
  \citenamefont {Goldman}, \citenamefont {Schmitt}, \citenamefont
  {Ghiringhelli}, \citenamefont {Barišić}, \citenamefont {Chan},
  \citenamefont {Dorow}, \citenamefont {Yu}, \citenamefont {Zhao},
  \citenamefont {Keimer},\ and\ \citenamefont {Greven}}]{tabis_charge_2014}%
  \BibitemOpen
  \bibfield  {author} {\bibinfo {author} {\bibfnamefont {W.}~\bibnamefont
  {Tabis}}, \bibinfo {author} {\bibfnamefont {Y.}~\bibnamefont {Li}}, \bibinfo
  {author} {\bibfnamefont {M.~L.}\ \bibnamefont {Tacon}}, \bibinfo {author}
  {\bibfnamefont {L.}~\bibnamefont {Braicovich}}, \bibinfo {author}
  {\bibfnamefont {A.}~\bibnamefont {Kreyssig}}, \bibinfo {author}
  {\bibfnamefont {M.}~\bibnamefont {Minola}}, \bibinfo {author} {\bibfnamefont
  {G.}~\bibnamefont {Dellea}}, \bibinfo {author} {\bibfnamefont
  {E.}~\bibnamefont {Weschke}}, \bibinfo {author} {\bibfnamefont {M.~J.}\
  \bibnamefont {Veit}}, \bibinfo {author} {\bibfnamefont {M.}~\bibnamefont
  {Ramazanoglu}}, \bibinfo {author} {\bibfnamefont {A.~I.}\ \bibnamefont
  {Goldman}}, \bibinfo {author} {\bibfnamefont {T.}~\bibnamefont {Schmitt}},
  \bibinfo {author} {\bibfnamefont {G.}~\bibnamefont {Ghiringhelli}}, \bibinfo
  {author} {\bibfnamefont {N.}~\bibnamefont {Barišić}}, \bibinfo {author}
  {\bibfnamefont {M.~K.}\ \bibnamefont {Chan}}, \bibinfo {author}
  {\bibfnamefont {C.~J.}\ \bibnamefont {Dorow}}, \bibinfo {author}
  {\bibfnamefont {G.}~\bibnamefont {Yu}}, \bibinfo {author} {\bibfnamefont
  {X.}~\bibnamefont {Zhao}}, \bibinfo {author} {\bibfnamefont {B.}~\bibnamefont
  {Keimer}},\ and\ \bibinfo {author} {\bibfnamefont {M.}~\bibnamefont
  {Greven}},\ }\bibfield  {title} {{\selectlanguage {english}\bibinfo {title}
  {Charge order and its connection with {Fermi}-liquid charge transport in a
  pristine high-{Tc} cuprate}},\ }\href {https://doi.org/10.1038/ncomms6875}
  {\bibfield  {journal} {\bibinfo  {journal} {Nature Communications}\ }\textbf
  {\bibinfo {volume} {5}},\ \bibinfo {pages} {5875} (\bibinfo {year}
  {2014})}\BibitemShut {NoStop}%
\bibitem [{\citenamefont {Da~Silva~Neto}\ \emph {et~al.}(2015)\citenamefont
  {Da~Silva~Neto}, \citenamefont {Comin}, \citenamefont {He}, \citenamefont
  {Sutarto}, \citenamefont {Jiang}, \citenamefont {Greene}, \citenamefont
  {Sawatzky},\ and\ \citenamefont {Damascelli}}]{da_silva_neto_charge_2015}%
  \BibitemOpen
  \bibfield  {author} {\bibinfo {author} {\bibfnamefont {E.~H.}\ \bibnamefont
  {Da~Silva~Neto}}, \bibinfo {author} {\bibfnamefont {R.}~\bibnamefont
  {Comin}}, \bibinfo {author} {\bibfnamefont {F.}~\bibnamefont {He}}, \bibinfo
  {author} {\bibfnamefont {R.}~\bibnamefont {Sutarto}}, \bibinfo {author}
  {\bibfnamefont {Y.}~\bibnamefont {Jiang}}, \bibinfo {author} {\bibfnamefont
  {R.~L.}\ \bibnamefont {Greene}}, \bibinfo {author} {\bibfnamefont {G.~A.}\
  \bibnamefont {Sawatzky}},\ and\ \bibinfo {author} {\bibfnamefont
  {A.}~\bibnamefont {Damascelli}},\ }\bibfield  {title} {{\selectlanguage
  {english}\bibinfo {title} {Charge ordering in the electron-doped
  superconductor {Nd2}–{xCexCuO4}}},\ }\href
  {https://doi.org/10.1126/science.1256441} {\bibfield  {journal} {\bibinfo
  {journal} {Science (American Association for the Advancement of Science)}\
  }\textbf {\bibinfo {volume} {347}},\ \bibinfo {pages} {282} (\bibinfo {year}
  {2015})},\ \bibinfo {note} {place: WASHINGTON Publisher: American Association
  for the Advancement of Science}\BibitemShut {NoStop}%
\bibitem [{\citenamefont {Peng}\ \emph {et~al.}(2016)\citenamefont {Peng},
  \citenamefont {Salluzzo}, \citenamefont {Sun}, \citenamefont {Ponti},
  \citenamefont {Betto}, \citenamefont {Ferretti}, \citenamefont {Fumagalli},
  \citenamefont {Kummer}, \citenamefont {Le~Tacon}, \citenamefont {Zhou},
  \citenamefont {Brookes}, \citenamefont {Braicovich},\ and\ \citenamefont
  {Ghiringhelli}}]{peng_direct_2016}%
  \BibitemOpen
  \bibfield  {author} {\bibinfo {author} {\bibfnamefont {Y.~Y.}\ \bibnamefont
  {Peng}}, \bibinfo {author} {\bibfnamefont {M.}~\bibnamefont {Salluzzo}},
  \bibinfo {author} {\bibfnamefont {X.}~\bibnamefont {Sun}}, \bibinfo {author}
  {\bibfnamefont {A.}~\bibnamefont {Ponti}}, \bibinfo {author} {\bibfnamefont
  {D.}~\bibnamefont {Betto}}, \bibinfo {author} {\bibfnamefont {A.~M.}\
  \bibnamefont {Ferretti}}, \bibinfo {author} {\bibfnamefont {F.}~\bibnamefont
  {Fumagalli}}, \bibinfo {author} {\bibfnamefont {K.}~\bibnamefont {Kummer}},
  \bibinfo {author} {\bibfnamefont {M.}~\bibnamefont {Le~Tacon}}, \bibinfo
  {author} {\bibfnamefont {X.~J.}\ \bibnamefont {Zhou}}, \bibinfo {author}
  {\bibfnamefont {N.~B.}\ \bibnamefont {Brookes}}, \bibinfo {author}
  {\bibfnamefont {L.}~\bibnamefont {Braicovich}},\ and\ \bibinfo {author}
  {\bibfnamefont {G.}~\bibnamefont {Ghiringhelli}},\ }\bibfield  {title}
  {{\selectlanguage {english}\bibinfo {title} {Direct observation of charge
  order in underdoped and optimally doped bi$_2$(sr,la)$_2$cuo$_{6+\delta}$ by
  resonant inelastic x-ray scattering}},\ }\href
  {https://doi.org/10.1103/PhysRevB.94.184511} {\bibfield  {journal} {\bibinfo
  {journal} {Physical Review B}\ }\textbf {\bibinfo {volume} {94}},\ \bibinfo
  {pages} {184511} (\bibinfo {year} {2016})}\BibitemShut {NoStop}%
\bibitem [{\citenamefont {Chang}\ \emph
  {et~al.}(2012{\natexlab{b}})\citenamefont {Chang}, \citenamefont {Blackburn},
  \citenamefont {Holmes}, \citenamefont {Christensen}, \citenamefont {Larsen},
  \citenamefont {Mesot}, \citenamefont {Liang}, \citenamefont {Bonn},
  \citenamefont {Hardy}, \citenamefont {Watenphul}, \citenamefont {von
  Zimmermann}, \citenamefont {Forgan},\ and\ \citenamefont
  {Hayden}}]{chang_direct_2012}%
  \BibitemOpen
  \bibfield  {author} {\bibinfo {author} {\bibfnamefont {J.}~\bibnamefont
  {Chang}}, \bibinfo {author} {\bibfnamefont {E.}~\bibnamefont {Blackburn}},
  \bibinfo {author} {\bibfnamefont {A.~T.}\ \bibnamefont {Holmes}}, \bibinfo
  {author} {\bibfnamefont {N.~B.}\ \bibnamefont {Christensen}}, \bibinfo
  {author} {\bibfnamefont {J.}~\bibnamefont {Larsen}}, \bibinfo {author}
  {\bibfnamefont {J.}~\bibnamefont {Mesot}}, \bibinfo {author} {\bibfnamefont
  {R.}~\bibnamefont {Liang}}, \bibinfo {author} {\bibfnamefont {D.~A.}\
  \bibnamefont {Bonn}}, \bibinfo {author} {\bibfnamefont {W.~N.}\ \bibnamefont
  {Hardy}}, \bibinfo {author} {\bibfnamefont {A.}~\bibnamefont {Watenphul}},
  \bibinfo {author} {\bibfnamefont {M.}~\bibnamefont {von Zimmermann}},
  \bibinfo {author} {\bibfnamefont {E.~M.}\ \bibnamefont {Forgan}},\ and\
  \bibinfo {author} {\bibfnamefont {S.~M.}\ \bibnamefont {Hayden}},\ }\bibfield
   {title} {{\selectlanguage {english}\bibinfo {title} {Direct observation of
  competition between superconductivity and charge density wave order in
  {YBa2Cu3O6}.67}},\ }\href {https://doi.org/10.1038/NPHYS2456} {\bibfield
  {journal} {\bibinfo  {journal} {Nature physics}\ }\textbf {\bibinfo {volume}
  {8}},\ \bibinfo {pages} {871} (\bibinfo {year} {2012}{\natexlab{b}})},\
  \bibinfo {note} {place: LONDON Publisher: NATURE PUBLISHING
  GROUP}\BibitemShut {NoStop}%
\bibitem [{\citenamefont {Kang}\ \emph {et~al.}(2019)\citenamefont {Kang},
  \citenamefont {Pelliciari}, \citenamefont {Frano}, \citenamefont {Breznay},
  \citenamefont {Schierle}, \citenamefont {Weschke}, \citenamefont {Sutarto},
  \citenamefont {He}, \citenamefont {Shafer}, \citenamefont {Arenholz},
  \citenamefont {Chen}, \citenamefont {Zhang}, \citenamefont {Ruiz},
  \citenamefont {Hao}, \citenamefont {Lewin}, \citenamefont {Analytis},
  \citenamefont {Krockenberger}, \citenamefont {Yamamoto}, \citenamefont
  {Das},\ and\ \citenamefont {Comin}}]{kang_evolution_2019}%
  \BibitemOpen
  \bibfield  {author} {\bibinfo {author} {\bibfnamefont {M.}~\bibnamefont
  {Kang}}, \bibinfo {author} {\bibfnamefont {J.}~\bibnamefont {Pelliciari}},
  \bibinfo {author} {\bibfnamefont {A.}~\bibnamefont {Frano}}, \bibinfo
  {author} {\bibfnamefont {N.}~\bibnamefont {Breznay}}, \bibinfo {author}
  {\bibfnamefont {E.}~\bibnamefont {Schierle}}, \bibinfo {author}
  {\bibfnamefont {E.}~\bibnamefont {Weschke}}, \bibinfo {author} {\bibfnamefont
  {R.}~\bibnamefont {Sutarto}}, \bibinfo {author} {\bibfnamefont
  {F.}~\bibnamefont {He}}, \bibinfo {author} {\bibfnamefont {P.}~\bibnamefont
  {Shafer}}, \bibinfo {author} {\bibfnamefont {E.}~\bibnamefont {Arenholz}},
  \bibinfo {author} {\bibfnamefont {M.}~\bibnamefont {Chen}}, \bibinfo {author}
  {\bibfnamefont {K.}~\bibnamefont {Zhang}}, \bibinfo {author} {\bibfnamefont
  {A.}~\bibnamefont {Ruiz}}, \bibinfo {author} {\bibfnamefont {Z.}~\bibnamefont
  {Hao}}, \bibinfo {author} {\bibfnamefont {S.}~\bibnamefont {Lewin}}, \bibinfo
  {author} {\bibfnamefont {J.}~\bibnamefont {Analytis}}, \bibinfo {author}
  {\bibfnamefont {Y.}~\bibnamefont {Krockenberger}}, \bibinfo {author}
  {\bibfnamefont {H.}~\bibnamefont {Yamamoto}}, \bibinfo {author}
  {\bibfnamefont {T.}~\bibnamefont {Das}},\ and\ \bibinfo {author}
  {\bibfnamefont {R.}~\bibnamefont {Comin}},\ }\bibfield  {title}
  {{\selectlanguage {english}\bibinfo {title} {Evolution of charge order
  topology across a magnetic phase transition in cuprate superconductors}},\
  }\href {https://doi.org/10.1038/s41567-018-0401-8} {\bibfield  {journal}
  {\bibinfo  {journal} {Nature Physics}\ }\textbf {\bibinfo {volume} {15}},\
  \bibinfo {pages} {335} (\bibinfo {year} {2019})}\BibitemShut {NoStop}%
\bibitem [{\citenamefont {Ghiringhelli}\ \emph
  {et~al.}(2012{\natexlab{b}})\citenamefont {Ghiringhelli}, \citenamefont
  {Le~Tacon}, \citenamefont {Minola}, \citenamefont {Blanco-Canosa},
  \citenamefont {Mazzoli}, \citenamefont {Brookes}, \citenamefont {De~Luca},
  \citenamefont {Frano}, \citenamefont {Hawthorn}, \citenamefont {He},
  \citenamefont {Loew}, \citenamefont {Moretti~Sala}, \citenamefont {Peets},
  \citenamefont {Salluzzo}, \citenamefont {Schierle}, \citenamefont {Sutarto},
  \citenamefont {Sawatzky}, \citenamefont {Weschke}, \citenamefont {Keimer},\
  and\ \citenamefont {Braicovich}}]{ghiringhelli_long-range_2012}%
  \BibitemOpen
  \bibfield  {author} {\bibinfo {author} {\bibfnamefont {G.}~\bibnamefont
  {Ghiringhelli}}, \bibinfo {author} {\bibfnamefont {M.}~\bibnamefont
  {Le~Tacon}}, \bibinfo {author} {\bibfnamefont {M.}~\bibnamefont {Minola}},
  \bibinfo {author} {\bibfnamefont {S.}~\bibnamefont {Blanco-Canosa}}, \bibinfo
  {author} {\bibfnamefont {C.}~\bibnamefont {Mazzoli}}, \bibinfo {author}
  {\bibfnamefont {N.~B.}\ \bibnamefont {Brookes}}, \bibinfo {author}
  {\bibfnamefont {G.~M.}\ \bibnamefont {De~Luca}}, \bibinfo {author}
  {\bibfnamefont {A.}~\bibnamefont {Frano}}, \bibinfo {author} {\bibfnamefont
  {D.~G.}\ \bibnamefont {Hawthorn}}, \bibinfo {author} {\bibfnamefont
  {F.}~\bibnamefont {He}}, \bibinfo {author} {\bibfnamefont {T.}~\bibnamefont
  {Loew}}, \bibinfo {author} {\bibfnamefont {M.}~\bibnamefont {Moretti~Sala}},
  \bibinfo {author} {\bibfnamefont {D.~C.}\ \bibnamefont {Peets}}, \bibinfo
  {author} {\bibfnamefont {M.}~\bibnamefont {Salluzzo}}, \bibinfo {author}
  {\bibfnamefont {E.}~\bibnamefont {Schierle}}, \bibinfo {author}
  {\bibfnamefont {R.}~\bibnamefont {Sutarto}}, \bibinfo {author} {\bibfnamefont
  {G.~A.}\ \bibnamefont {Sawatzky}}, \bibinfo {author} {\bibfnamefont
  {E.}~\bibnamefont {Weschke}}, \bibinfo {author} {\bibfnamefont
  {B.}~\bibnamefont {Keimer}},\ and\ \bibinfo {author} {\bibfnamefont
  {L.}~\bibnamefont {Braicovich}},\ }\bibfield  {title} {{\selectlanguage
  {english}\bibinfo {title} {Long-range incommensurate charge fluctuations in
  ({Y},{Nd}){Ba} {2Cu3O6}+x}},\ }\href
  {https://doi.org/10.1126/science.1223532} {\bibfield  {journal} {\bibinfo
  {journal} {Science (American Association for the Advancement of Science)}\
  }\textbf {\bibinfo {volume} {337}},\ \bibinfo {pages} {821} (\bibinfo {year}
  {2012}{\natexlab{b}})}\BibitemShut {NoStop}%
\bibitem [{\citenamefont {Li}\ \emph {et~al.}(2020)\citenamefont {Li},
  \citenamefont {Nag}, \citenamefont {Pelliciari}, \citenamefont {Robarts},
  \citenamefont {Walters}, \citenamefont {Garcia-Fernandez}, \citenamefont
  {Eisaki}, \citenamefont {Song}, \citenamefont {Ding}, \citenamefont
  {Johnston}, \citenamefont {Comin},\ and\ \citenamefont
  {Zhou}}]{li_multiorbital_2020}%
  \BibitemOpen
  \bibfield  {author} {\bibinfo {author} {\bibfnamefont {J.}~\bibnamefont
  {Li}}, \bibinfo {author} {\bibfnamefont {A.}~\bibnamefont {Nag}}, \bibinfo
  {author} {\bibfnamefont {J.}~\bibnamefont {Pelliciari}}, \bibinfo {author}
  {\bibfnamefont {H.}~\bibnamefont {Robarts}}, \bibinfo {author} {\bibfnamefont
  {A.}~\bibnamefont {Walters}}, \bibinfo {author} {\bibfnamefont
  {M.}~\bibnamefont {Garcia-Fernandez}}, \bibinfo {author} {\bibfnamefont
  {H.}~\bibnamefont {Eisaki}}, \bibinfo {author} {\bibfnamefont
  {D.}~\bibnamefont {Song}}, \bibinfo {author} {\bibfnamefont {H.}~\bibnamefont
  {Ding}}, \bibinfo {author} {\bibfnamefont {S.}~\bibnamefont {Johnston}},
  \bibinfo {author} {\bibfnamefont {R.}~\bibnamefont {Comin}},\ and\ \bibinfo
  {author} {\bibfnamefont {K.-J.}\ \bibnamefont {Zhou}},\ }\bibfield  {title}
  {{\selectlanguage {english}\bibinfo {title} {Multiorbital charge-density wave
  excitations and concomitant phonon anomalies in
  bi$_2$sr$_2$lacuo$_{6+\delta}$}},\ }\href
  {https://doi.org/10.1073/pnas.2001755117} {\bibfield  {journal} {\bibinfo
  {journal} {Proceedings of the National Academy of Sciences}\ }\textbf
  {\bibinfo {volume} {117}},\ \bibinfo {pages} {16219} (\bibinfo {year}
  {2020})}\BibitemShut {NoStop}%
\bibitem [{\citenamefont {Abbamonte}\ \emph {et~al.}(2005)\citenamefont
  {Abbamonte}, \citenamefont {Rusydi}, \citenamefont {Smadici}, \citenamefont
  {Gu}, \citenamefont {Sawatzky},\ and\ \citenamefont
  {Feng}}]{abbamonte_spatially_2005}%
  \BibitemOpen
  \bibfield  {author} {\bibinfo {author} {\bibfnamefont {P.}~\bibnamefont
  {Abbamonte}}, \bibinfo {author} {\bibfnamefont {A.}~\bibnamefont {Rusydi}},
  \bibinfo {author} {\bibfnamefont {S.}~\bibnamefont {Smadici}}, \bibinfo
  {author} {\bibfnamefont {G.~D.}\ \bibnamefont {Gu}}, \bibinfo {author}
  {\bibfnamefont {G.~A.}\ \bibnamefont {Sawatzky}},\ and\ \bibinfo {author}
  {\bibfnamefont {D.~L.}\ \bibnamefont {Feng}},\ }\bibfield  {title}
  {{\selectlanguage {english}\bibinfo {title} {Spatially modulated '{Mottness}'
  in la$_{2-x}$ba$_x$cuo$_4$}},\ }\href {https://doi.org/10.1038/nphys178}
  {\bibfield  {journal} {\bibinfo  {journal} {Nature physics}\ }\textbf
  {\bibinfo {volume} {1}},\ \bibinfo {pages} {155} (\bibinfo {year} {2005})},\
  \bibinfo {note} {place: LONDON Publisher: NATURE PUBLISHING
  GROUP}\BibitemShut {NoStop}%
\bibitem [{\citenamefont {Comin}\ \emph
  {et~al.}(2015{\natexlab{a}})\citenamefont {Comin}, \citenamefont {Sutarto},
  \citenamefont {He}, \citenamefont {Da~Silva~Neto}, \citenamefont {Chauviere},
  \citenamefont {Fraño}, \citenamefont {Liang}, \citenamefont {Hardy},
  \citenamefont {Bonn}, \citenamefont {Yoshida}, \citenamefont {Eisaki},
  \citenamefont {Achkar}, \citenamefont {Hawthorn}, \citenamefont {Keimer},
  \citenamefont {Sawatzky},\ and\ \citenamefont
  {Damascelli}}]{comin_symmetry_2015}%
  \BibitemOpen
  \bibfield  {author} {\bibinfo {author} {\bibfnamefont {R.}~\bibnamefont
  {Comin}}, \bibinfo {author} {\bibfnamefont {R.}~\bibnamefont {Sutarto}},
  \bibinfo {author} {\bibfnamefont {F.}~\bibnamefont {He}}, \bibinfo {author}
  {\bibfnamefont {E.~H.}\ \bibnamefont {Da~Silva~Neto}}, \bibinfo {author}
  {\bibfnamefont {L.}~\bibnamefont {Chauviere}}, \bibinfo {author}
  {\bibfnamefont {A.}~\bibnamefont {Fraño}}, \bibinfo {author} {\bibfnamefont
  {R.}~\bibnamefont {Liang}}, \bibinfo {author} {\bibfnamefont {W.~N.}\
  \bibnamefont {Hardy}}, \bibinfo {author} {\bibfnamefont {D.~A.}\ \bibnamefont
  {Bonn}}, \bibinfo {author} {\bibfnamefont {Y.}~\bibnamefont {Yoshida}},
  \bibinfo {author} {\bibfnamefont {H.}~\bibnamefont {Eisaki}}, \bibinfo
  {author} {\bibfnamefont {A.~J.}\ \bibnamefont {Achkar}}, \bibinfo {author}
  {\bibfnamefont {D.~G.}\ \bibnamefont {Hawthorn}}, \bibinfo {author}
  {\bibfnamefont {B.}~\bibnamefont {Keimer}}, \bibinfo {author} {\bibfnamefont
  {G.~A.}\ \bibnamefont {Sawatzky}},\ and\ \bibinfo {author} {\bibfnamefont
  {A.}~\bibnamefont {Damascelli}},\ }\bibfield  {title} {{\selectlanguage
  {english}\bibinfo {title} {Symmetry of charge order in cuprates}},\ }\href
  {https://doi.org/10.1038/nmat4295} {\bibfield  {journal} {\bibinfo  {journal}
  {Nature materials}\ }\textbf {\bibinfo {volume} {14}},\ \bibinfo {pages}
  {796} (\bibinfo {year} {2015}{\natexlab{a}})},\ \bibinfo {note} {place:
  LONDON Publisher: NATURE PUBLISHING GROUP}\BibitemShut {NoStop}%
\bibitem [{\citenamefont {da~Silva~Neto}\ \emph {et~al.}(2014)\citenamefont
  {da~Silva~Neto}, \citenamefont {Aynajian}, \citenamefont {Frano},
  \citenamefont {Comin}, \citenamefont {Schierle}, \citenamefont {Weschke},
  \citenamefont {Gyenis}, \citenamefont {Wen}, \citenamefont {Schneeloch},
  \citenamefont {Xu}, \citenamefont {Ono}, \citenamefont {Gu}, \citenamefont
  {Le~Tacon},\ and\ \citenamefont {Yazdani}}]{da_silva_neto_ubiquitous_2014}%
  \BibitemOpen
  \bibfield  {author} {\bibinfo {author} {\bibfnamefont {E.~H.}\ \bibnamefont
  {da~Silva~Neto}}, \bibinfo {author} {\bibfnamefont {P.}~\bibnamefont
  {Aynajian}}, \bibinfo {author} {\bibfnamefont {A.}~\bibnamefont {Frano}},
  \bibinfo {author} {\bibfnamefont {R.}~\bibnamefont {Comin}}, \bibinfo
  {author} {\bibfnamefont {E.}~\bibnamefont {Schierle}}, \bibinfo {author}
  {\bibfnamefont {E.}~\bibnamefont {Weschke}}, \bibinfo {author} {\bibfnamefont
  {A.}~\bibnamefont {Gyenis}}, \bibinfo {author} {\bibfnamefont
  {J.}~\bibnamefont {Wen}}, \bibinfo {author} {\bibfnamefont {J.}~\bibnamefont
  {Schneeloch}}, \bibinfo {author} {\bibfnamefont {Z.}~\bibnamefont {Xu}},
  \bibinfo {author} {\bibfnamefont {S.}~\bibnamefont {Ono}}, \bibinfo {author}
  {\bibfnamefont {G.}~\bibnamefont {Gu}}, \bibinfo {author} {\bibfnamefont
  {M.}~\bibnamefont {Le~Tacon}},\ and\ \bibinfo {author} {\bibfnamefont
  {A.}~\bibnamefont {Yazdani}},\ }\bibfield  {title} {{\selectlanguage
  {english}\bibinfo {title} {Ubiquitous {Interplay} {Between} {Charge}
  {Ordering} and {High}-{Temperature} {Superconductivity} in {Cuprates}}},\
  }\href {https://doi.org/10.1126/science.1243479} {\bibfield  {journal}
  {\bibinfo  {journal} {Science}\ }\textbf {\bibinfo {volume} {343}},\ \bibinfo
  {pages} {393} (\bibinfo {year} {2014})}\BibitemShut {NoStop}%
\bibitem [{\citenamefont {Phillabaum}\ \emph {et~al.}(2012)\citenamefont
  {Phillabaum}, \citenamefont {Carlson},\ and\ \citenamefont
  {Dahmen}}]{PhillabaumNatComm2012}%
  \BibitemOpen
  \bibfield  {author} {\bibinfo {author} {\bibfnamefont {B.}~\bibnamefont
  {Phillabaum}}, \bibinfo {author} {\bibfnamefont {E.~W.}\ \bibnamefont
  {Carlson}},\ and\ \bibinfo {author} {\bibfnamefont {K.~A.}\ \bibnamefont
  {Dahmen}},\ }\bibfield  {title} {\bibinfo {title} {Spatial complexity due to
  bulk electronic nematicity in a superconducting underdoped cuprate},\ }\href
  {https://doi.org/10.1038/ncomms1920} {\bibfield  {journal} {\bibinfo
  {journal} {Nature Communications}\ }\textbf {\bibinfo {volume} {3}},\
  \bibinfo {pages} {915} (\bibinfo {year} {2012})}\BibitemShut {NoStop}%
\bibitem [{\citenamefont {Lawler}\ \emph {et~al.}(2010)\citenamefont {Lawler},
  \citenamefont {Fujita}, \citenamefont {Lee}, \citenamefont {Schmidt},
  \citenamefont {Kohsaka}, \citenamefont {Kim}, \citenamefont {Eisaki},
  \citenamefont {Uchida}, \citenamefont {Davis}, \citenamefont {Sethna},\ and\
  \citenamefont {Kim}}]{LawlerNature2010}%
  \BibitemOpen
  \bibfield  {author} {\bibinfo {author} {\bibfnamefont {M.~J.}\ \bibnamefont
  {Lawler}}, \bibinfo {author} {\bibfnamefont {K.}~\bibnamefont {Fujita}},
  \bibinfo {author} {\bibfnamefont {J.}~\bibnamefont {Lee}}, \bibinfo {author}
  {\bibfnamefont {A.~R.}\ \bibnamefont {Schmidt}}, \bibinfo {author}
  {\bibfnamefont {Y.}~\bibnamefont {Kohsaka}}, \bibinfo {author} {\bibfnamefont
  {C.~K.}\ \bibnamefont {Kim}}, \bibinfo {author} {\bibfnamefont
  {H.}~\bibnamefont {Eisaki}}, \bibinfo {author} {\bibfnamefont
  {S.}~\bibnamefont {Uchida}}, \bibinfo {author} {\bibfnamefont {J.~C.}\
  \bibnamefont {Davis}}, \bibinfo {author} {\bibfnamefont {J.~P.}\ \bibnamefont
  {Sethna}},\ and\ \bibinfo {author} {\bibfnamefont {E.-A.}\ \bibnamefont
  {Kim}},\ }\bibfield  {title} {\bibinfo {title} {Intra-unit-cell electronic
  nematicity of the high-$t_c$ copper-oxide pseudogap states},\ }\href
  {https://doi.org/10.1038/nature09169} {\bibfield  {journal} {\bibinfo
  {journal} {Nature}\ }\textbf {\bibinfo {volume} {466}},\ \bibinfo {pages}
  {347} (\bibinfo {year} {2010})}\BibitemShut {NoStop}%
\bibitem [{\citenamefont {Robertson}\ \emph {et~al.}(2006)\citenamefont
  {Robertson}, \citenamefont {Kivelson}, \citenamefont {Fradkin}, \citenamefont
  {Fang},\ and\ \citenamefont {Kapitulnik}}]{RobertsonPRB2006}%
  \BibitemOpen
  \bibfield  {author} {\bibinfo {author} {\bibfnamefont {J.}~\bibnamefont
  {Robertson}}, \bibinfo {author} {\bibfnamefont {S.}~\bibnamefont {Kivelson}},
  \bibinfo {author} {\bibfnamefont {E.}~\bibnamefont {Fradkin}}, \bibinfo
  {author} {\bibfnamefont {A.}~\bibnamefont {Fang}},\ and\ \bibinfo {author}
  {\bibfnamefont {A.}~\bibnamefont {Kapitulnik}},\ }\bibfield  {title}
  {\bibinfo {title} {Distinguishing patterns of charge order: Stripes or
  checkerboards},\ }\href {https://doi.org/10.1103/PhysRevB.74.134507}
  {\bibfield  {journal} {\bibinfo  {journal} {Physical Review B}\ }\textbf
  {\bibinfo {volume} {74}},\ \bibinfo {pages} {134507} (\bibinfo {year}
  {2006})}\BibitemShut {NoStop}%
\bibitem [{\citenamefont {{Del Maestro}}\ \emph {et~al.}(2006)\citenamefont
  {{Del Maestro}}, \citenamefont {Rosenow},\ and\ \citenamefont
  {Sachdev}}]{DelMaestroPRB2006}%
  \BibitemOpen
  \bibfield  {author} {\bibinfo {author} {\bibfnamefont {A.}~\bibnamefont {{Del
  Maestro}}}, \bibinfo {author} {\bibfnamefont {B.}~\bibnamefont {Rosenow}},\
  and\ \bibinfo {author} {\bibfnamefont {S.}~\bibnamefont {Sachdev}},\
  }\bibfield  {title} {\bibinfo {title} {From stripe to checkerboard ordering
  of charge-density waves on the square lattice in the presence of quenched
  disorder},\ }\href {https://doi.org/10.1103/PhysRevB.74.024520} {\bibfield
  {journal} {\bibinfo  {journal} {Physical Review B}\ }\textbf {\bibinfo
  {volume} {74}},\ \bibinfo {pages} {024520} (\bibinfo {year}
  {2006})}\BibitemShut {NoStop}%
\bibitem [{\citenamefont {Parker}\ \emph {et~al.}(2010)\citenamefont {Parker},
  \citenamefont {Aynajian}, \citenamefont {{da Silva Neto}}, \citenamefont
  {Pushp}, \citenamefont {Ono}, \citenamefont {Wen}, \citenamefont {Xu},
  \citenamefont {Gu},\ and\ \citenamefont {Yazdani}}]{ParkerNature2010}%
  \BibitemOpen
  \bibfield  {author} {\bibinfo {author} {\bibfnamefont {C.~V.}\ \bibnamefont
  {Parker}}, \bibinfo {author} {\bibfnamefont {P.}~\bibnamefont {Aynajian}},
  \bibinfo {author} {\bibfnamefont {E.~H.}\ \bibnamefont {{da Silva Neto}}},
  \bibinfo {author} {\bibfnamefont {A.}~\bibnamefont {Pushp}}, \bibinfo
  {author} {\bibfnamefont {S.}~\bibnamefont {Ono}}, \bibinfo {author}
  {\bibfnamefont {J.}~\bibnamefont {Wen}}, \bibinfo {author} {\bibfnamefont
  {Z.}~\bibnamefont {Xu}}, \bibinfo {author} {\bibfnamefont {G.}~\bibnamefont
  {Gu}},\ and\ \bibinfo {author} {\bibfnamefont {A.}~\bibnamefont {Yazdani}},\
  }\bibfield  {title} {\bibinfo {title} {Fluctuating stripes at the onset of
  the pseudogap in the high-$t_c$ superconductor
  {Bi$_2$Sr$_2$CaCu$_2$O$_{8+d}$}},\ }\href
  {https://doi.org/10.1038/nature09597} {\bibfield  {journal} {\bibinfo
  {journal} {Nature}\ }\textbf {\bibinfo {volume} {468}},\ \bibinfo {pages}
  {677} (\bibinfo {year} {2010})}\BibitemShut {NoStop}%
\bibitem [{\citenamefont {Fujita}\ \emph {et~al.}(2014)\citenamefont {Fujita},
  \citenamefont {Hamidian}, \citenamefont {Edkins}, \citenamefont {Kim},
  \citenamefont {Kohsaka}, \citenamefont {Azuma}, \citenamefont {Takano},
  \citenamefont {Takagi}, \citenamefont {Eisaki}, \citenamefont {Uchida},
  \citenamefont {Allais}, \citenamefont {Lawler}, \citenamefont {Kim},
  \citenamefont {Sachdev},\ and\ \citenamefont {Davis}}]{FujitaPNAS2014}%
  \BibitemOpen
  \bibfield  {author} {\bibinfo {author} {\bibfnamefont {K.}~\bibnamefont
  {Fujita}}, \bibinfo {author} {\bibfnamefont {M.~H.}\ \bibnamefont
  {Hamidian}}, \bibinfo {author} {\bibfnamefont {S.~D.}\ \bibnamefont
  {Edkins}}, \bibinfo {author} {\bibfnamefont {C.~K.}\ \bibnamefont {Kim}},
  \bibinfo {author} {\bibfnamefont {Y.}~\bibnamefont {Kohsaka}}, \bibinfo
  {author} {\bibfnamefont {M.}~\bibnamefont {Azuma}}, \bibinfo {author}
  {\bibfnamefont {M.}~\bibnamefont {Takano}}, \bibinfo {author} {\bibfnamefont
  {H.}~\bibnamefont {Takagi}}, \bibinfo {author} {\bibfnamefont
  {H.}~\bibnamefont {Eisaki}}, \bibinfo {author} {\bibfnamefont {S.-i.}\
  \bibnamefont {Uchida}}, \bibinfo {author} {\bibfnamefont {A.}~\bibnamefont
  {Allais}}, \bibinfo {author} {\bibfnamefont {M.~J.}\ \bibnamefont {Lawler}},
  \bibinfo {author} {\bibfnamefont {E.-a.}\ \bibnamefont {Kim}}, \bibinfo
  {author} {\bibfnamefont {S.}~\bibnamefont {Sachdev}},\ and\ \bibinfo {author}
  {\bibfnamefont {J.~C.~S.}\ \bibnamefont {Davis}},\ }\bibfield  {title}
  {\bibinfo {title} {Direct phase-sensitive identification of a $d$-form factor
  density wave in underdoped cuprates},\ }\href
  {https://doi.org/10.1073/pnas.1406297111} {\bibfield  {journal} {\bibinfo
  {journal} {Proceedings of the National Academy of Sciences}\ }\textbf
  {\bibinfo {volume} {111}},\ \bibinfo {pages} {E3026} (\bibinfo {year}
  {2014})}\BibitemShut {NoStop}%
\bibitem [{\citenamefont {Tranquada}\ \emph {et~al.}(1995)\citenamefont
  {Tranquada}, \citenamefont {Sternlieb}, \citenamefont {Axe}, \citenamefont
  {Nakamura},\ and\ \citenamefont {Uchida}}]{TranquadaNature1995}%
  \BibitemOpen
  \bibfield  {author} {\bibinfo {author} {\bibfnamefont {J.~M.}\ \bibnamefont
  {Tranquada}}, \bibinfo {author} {\bibfnamefont {B.~J.}\ \bibnamefont
  {Sternlieb}}, \bibinfo {author} {\bibfnamefont {J.~D.}\ \bibnamefont {Axe}},
  \bibinfo {author} {\bibfnamefont {Y.}~\bibnamefont {Nakamura}},\ and\
  \bibinfo {author} {\bibfnamefont {S.}~\bibnamefont {Uchida}},\ }\bibfield
  {title} {\bibinfo {title} {Evidence for stripe correlations of spins and
  holes in copper oxide superconductors},\ }\href
  {https://doi.org/10.1038/375561a0} {\bibfield  {journal} {\bibinfo  {journal}
  {Nature}\ }\textbf {\bibinfo {volume} {375}},\ \bibinfo {pages} {561}
  (\bibinfo {year} {1995})}\BibitemShut {NoStop}%
\bibitem [{\citenamefont {Mook}\ \emph {et~al.}(1998)\citenamefont {Mook},
  \citenamefont {Dai},\ and\ \citenamefont {Dog}}]{MookNature1998}%
  \BibitemOpen
  \bibfield  {author} {\bibinfo {author} {\bibfnamefont {H.~A.}\ \bibnamefont
  {Mook}}, \bibinfo {author} {\bibfnamefont {P.}~\bibnamefont {Dai}},\ and\
  \bibinfo {author} {\bibfnamefont {F.}~\bibnamefont {Dog}},\ }\bibfield
  {title} {\bibinfo {title} {Spin fluctuations in {YBa$_2$Cu$_3$O$_{6.6}$}},\
  }\href {https://doi.org/10.1038/26931} {\bibfield  {journal} {\bibinfo
  {journal} {Nature}\ }\textbf {\bibinfo {volume} {395}},\ \bibinfo {pages}
  {580} (\bibinfo {year} {1998})}\BibitemShut {NoStop}%
\bibitem [{\citenamefont {Comin}\ \emph
  {et~al.}(2015{\natexlab{b}})\citenamefont {Comin}, \citenamefont {Sutarto},
  \citenamefont {da~Silva~Neto}, \citenamefont {Chauviere}, \citenamefont
  {Liang}, \citenamefont {Hardy}, \citenamefont {Bonn}, \citenamefont {He},
  \citenamefont {Sawatzky},\ and\ \citenamefont
  {Damascelli}}]{CominScience2015}%
  \BibitemOpen
  \bibfield  {author} {\bibinfo {author} {\bibfnamefont {R.}~\bibnamefont
  {Comin}}, \bibinfo {author} {\bibfnamefont {R.}~\bibnamefont {Sutarto}},
  \bibinfo {author} {\bibfnamefont {E.~H.}\ \bibnamefont {da~Silva~Neto}},
  \bibinfo {author} {\bibfnamefont {L.}~\bibnamefont {Chauviere}}, \bibinfo
  {author} {\bibfnamefont {R.}~\bibnamefont {Liang}}, \bibinfo {author}
  {\bibfnamefont {W.~N.}\ \bibnamefont {Hardy}}, \bibinfo {author}
  {\bibfnamefont {D.~A.}\ \bibnamefont {Bonn}}, \bibinfo {author}
  {\bibfnamefont {F.}~\bibnamefont {He}}, \bibinfo {author} {\bibfnamefont
  {G.~A.}\ \bibnamefont {Sawatzky}},\ and\ \bibinfo {author} {\bibfnamefont
  {A.}~\bibnamefont {Damascelli}},\ }\bibfield  {title} {\bibinfo {title}
  {Broken translational and rotational symmetry via charge stripe order in
  underdoped {YBa$_2$Cu$_3$O$_{6+y}$}},\ }\href
  {https://doi.org/10.1126/science.1258399} {\bibfield  {journal} {\bibinfo
  {journal} {Science}\ }\textbf {\bibinfo {volume} {347}},\ \bibinfo {pages}
  {1335} (\bibinfo {year} {2015}{\natexlab{b}})}\BibitemShut {NoStop}%
\bibitem [{\citenamefont {Howald}\ \emph
  {et~al.}(2003{\natexlab{b}})\citenamefont {Howald}, \citenamefont {Eisaki},
  \citenamefont {Kaneko}, \citenamefont {Greven},\ and\ \citenamefont
  {Kapitulnik}}]{HowaldPRB2003}%
  \BibitemOpen
  \bibfield  {author} {\bibinfo {author} {\bibfnamefont {C.}~\bibnamefont
  {Howald}}, \bibinfo {author} {\bibfnamefont {H.}~\bibnamefont {Eisaki}},
  \bibinfo {author} {\bibfnamefont {N.}~\bibnamefont {Kaneko}}, \bibinfo
  {author} {\bibfnamefont {M.}~\bibnamefont {Greven}},\ and\ \bibinfo {author}
  {\bibfnamefont {A.}~\bibnamefont {Kapitulnik}},\ }\bibfield  {title}
  {\bibinfo {title} {Periodic density-of-states modulations in superconducting
  ${\mathrm{bi}}_{2}{\mathrm{sr}}_{2}{\mathrm{cacu}}_{2}{\mathrm{o}}_{8+\ensuremath{\delta}}$},\
  }\href {https://doi.org/10.1103/PhysRevB.67.014533} {\bibfield  {journal}
  {\bibinfo  {journal} {Phys. Rev. B}\ }\textbf {\bibinfo {volume} {67}},\
  \bibinfo {pages} {014533} (\bibinfo {year} {2003}{\natexlab{b}})}\BibitemShut
  {NoStop}%
\bibitem [{\citenamefont {Vershinin}\ \emph {et~al.}(2004)\citenamefont
  {Vershinin}, \citenamefont {Misra}, \citenamefont {Ono}, \citenamefont {Abe},
  \citenamefont {Ando},\ and\ \citenamefont {Yazdani}}]{VershininScience1995}%
  \BibitemOpen
  \bibfield  {author} {\bibinfo {author} {\bibfnamefont {M.}~\bibnamefont
  {Vershinin}}, \bibinfo {author} {\bibfnamefont {S.}~\bibnamefont {Misra}},
  \bibinfo {author} {\bibfnamefont {S.}~\bibnamefont {Ono}}, \bibinfo {author}
  {\bibfnamefont {Y.}~\bibnamefont {Abe}}, \bibinfo {author} {\bibfnamefont
  {Y.}~\bibnamefont {Ando}},\ and\ \bibinfo {author} {\bibfnamefont
  {A.}~\bibnamefont {Yazdani}},\ }\bibfield  {title} {\bibinfo {title} {Local
  ordering in the pseudogap state of the high-tc superconductor
  bi$_2$sr$_2$cacu$_2$o$_{8+\delta}$},\ }\href
  {https://doi.org/10.1126/science.1093384} {\bibfield  {journal} {\bibinfo
  {journal} {Science}\ }\textbf {\bibinfo {volume} {303}},\ \bibinfo {pages}
  {1995} (\bibinfo {year} {2004})},\ \Eprint
  {https://arxiv.org/abs/https://science.sciencemag.org/content/303/5666/1995.full.pdf}
  {https://science.sciencemag.org/content/303/5666/1995.full.pdf} \BibitemShut
  {NoStop}%
\bibitem [{\citenamefont {Arpaia}\ \emph {et~al.}(2019)\citenamefont {Arpaia},
  \citenamefont {Caprara}, \citenamefont {Fumagalli}, \citenamefont
  {De~Vecchi}, \citenamefont {Peng}, \citenamefont {Andersson}, \citenamefont
  {Betto}, \citenamefont {De~Luca}, \citenamefont {Brookes}, \citenamefont
  {Lombardi}, \citenamefont {Salluzzo}, \citenamefont {Braicovich},
  \citenamefont {Di~Castro}, \citenamefont {Grilli},\ and\ \citenamefont
  {Ghiringhelli}}]{ArpaiaScience2020}%
  \BibitemOpen
  \bibfield  {author} {\bibinfo {author} {\bibfnamefont {R.}~\bibnamefont
  {Arpaia}}, \bibinfo {author} {\bibfnamefont {S.}~\bibnamefont {Caprara}},
  \bibinfo {author} {\bibfnamefont {R.}~\bibnamefont {Fumagalli}}, \bibinfo
  {author} {\bibfnamefont {G.}~\bibnamefont {De~Vecchi}}, \bibinfo {author}
  {\bibfnamefont {Y.~Y.}\ \bibnamefont {Peng}}, \bibinfo {author}
  {\bibfnamefont {E.}~\bibnamefont {Andersson}}, \bibinfo {author}
  {\bibfnamefont {D.}~\bibnamefont {Betto}}, \bibinfo {author} {\bibfnamefont
  {G.~M.}\ \bibnamefont {De~Luca}}, \bibinfo {author} {\bibfnamefont {N.~B.}\
  \bibnamefont {Brookes}}, \bibinfo {author} {\bibfnamefont {F.}~\bibnamefont
  {Lombardi}}, \bibinfo {author} {\bibfnamefont {M.}~\bibnamefont {Salluzzo}},
  \bibinfo {author} {\bibfnamefont {L.}~\bibnamefont {Braicovich}}, \bibinfo
  {author} {\bibfnamefont {C.}~\bibnamefont {Di~Castro}}, \bibinfo {author}
  {\bibfnamefont {M.}~\bibnamefont {Grilli}},\ and\ \bibinfo {author}
  {\bibfnamefont {G.}~\bibnamefont {Ghiringhelli}},\ }\bibfield  {title}
  {\bibinfo {title} {Dynamical charge density fluctuations pervading the phase
  diagram of a cu-based high-tc superconductor},\ }\href
  {https://doi.org/10.1126/science.aav1315} {\bibfield  {journal} {\bibinfo
  {journal} {Science}\ }\textbf {\bibinfo {volume} {365}},\ \bibinfo {pages}
  {906} (\bibinfo {year} {2019})},\ \Eprint
  {https://arxiv.org/abs/https://science.sciencemag.org/content/365/6456/906.full.pdf}
  {https://science.sciencemag.org/content/365/6456/906.full.pdf} \BibitemShut
  {NoStop}%
\bibitem [{Sup()}]{Supplement}%
  \BibitemOpen
  \href@noop {} {}\bibinfo {note} {See Supplementary Information.}\BibitemShut
  {Stop}%
\bibitem [{\citenamefont {Fisher}(1967)}]{Fisher1967}%
  \BibitemOpen
  \bibfield  {author} {\bibinfo {author} {\bibfnamefont {M.~E.}\ \bibnamefont
  {Fisher}},\ }\bibfield  {title} {\bibinfo {title} {The theory of condensation
  and the critical point},\ }\href
  {https://doi.org/10.1103/PhysicsPhysiqueFizika.3.255} {\bibfield  {journal}
  {\bibinfo  {journal} {Physics Physique Fizika}\ }\textbf {\bibinfo {volume}
  {3}},\ \bibinfo {pages} {255} (\bibinfo {year} {1967})}\BibitemShut {NoStop}%
\bibitem [{\citenamefont {Newman}(2005)}]{NewmanContPhys2005}%
  \BibitemOpen
  \bibfield  {author} {\bibinfo {author} {\bibfnamefont {M.~E.~J.}\
  \bibnamefont {Newman}},\ }\bibfield  {title} {\bibinfo {title} {Power laws,
  {P}areto distributions and {Z}ipf's law},\ }\href
  {https://doi.org/10.1080/00107510500052444} {\bibfield  {journal} {\bibinfo
  {journal} {Contemporary Physics}\ }\textbf {\bibinfo {volume} {46}},\
  \bibinfo {pages} {323} (\bibinfo {year} {2005})}\BibitemShut {NoStop}%
\bibitem [{\citenamefont {Stauffer}\ and\ \citenamefont
  {Aharony}(1994)}]{StaufferBook1994}%
  \BibitemOpen
  \bibfield  {author} {\bibinfo {author} {\bibfnamefont {D.}~\bibnamefont
  {Stauffer}}\ and\ \bibinfo {author} {\bibfnamefont {A.}~\bibnamefont
  {Aharony}},\ }\href@noop {} {\emph {\bibinfo {title} {Introduction to
  percolation theory}}}\ (\bibinfo  {publisher} {CRC Press},\ \bibinfo {year}
  {1994})\BibitemShut {NoStop}%
\bibitem [{\citenamefont {Liu}\ \emph {et~al.}(2021)\citenamefont {Liu},
  \citenamefont {Carlson},\ and\ \citenamefont {Dahmen}}]{C-3Dx}%
  \BibitemOpen
  \bibfield  {author} {\bibinfo {author} {\bibfnamefont {S.}~\bibnamefont
  {Liu}}, \bibinfo {author} {\bibfnamefont {E.~W.}\ \bibnamefont {Carlson}},\
  and\ \bibinfo {author} {\bibfnamefont {K.~A.}\ \bibnamefont {Dahmen}},\
  }\bibfield  {title} {\bibinfo {title} {{Connecting Complex Electronic Pattern
  Formation to Critical Exponents}},\ }\href
  {https://doi.org/10.3390/condmat6040039} {\bibfield  {journal} {\bibinfo
  {journal} {Condensed Matter}\ }\textbf {\bibinfo {volume} {6}},\ \bibinfo
  {pages} {39} (\bibinfo {year} {2021})}\BibitemShut {NoStop}%
\bibitem [{\citenamefont {Carlson}\ \emph {et~al.}(2004)\citenamefont
  {Carlson}, \citenamefont {Emery}, \citenamefont {Kivelson},\ and\
  \citenamefont {Orgad}}]{concepts}%
  \BibitemOpen
  \bibfield  {author} {\bibinfo {author} {\bibfnamefont {E.~W.}\ \bibnamefont
  {Carlson}}, \bibinfo {author} {\bibfnamefont {V.~J.}\ \bibnamefont {Emery}},
  \bibinfo {author} {\bibfnamefont {S.~A.}\ \bibnamefont {Kivelson}},\ and\
  \bibinfo {author} {\bibfnamefont {D.}~\bibnamefont {Orgad}},\ }\bibinfo
  {title} {Concepts in high temperature superconductivity},\ in\ \href
  {http://xxx.lanl.gov/abs/cond-mat/0206217} {\emph {\bibinfo {booktitle} {The
  Physics of Superconductors, Vol. II}}},\ \bibinfo {editor} {edited by\
  \bibinfo {editor} {\bibfnamefont {J.}~\bibnamefont {Ketterson}}\ and\
  \bibinfo {editor} {\bibfnamefont {K.}~\bibnamefont {Benneman}}}\ (\bibinfo
  {publisher} {Springer-Verlag},\ \bibinfo {year} {2004})\ \bibinfo {note} {in
  The Physics of Superconductors, Vol. II, ed. J.~Ketterson and
  K.~Benneman}\BibitemShut {NoStop}%
\bibitem [{\citenamefont {Emery}\ \emph {et~al.}(1997)\citenamefont {Emery},
  \citenamefont {Kivelson},\ and\ \citenamefont {Zachar}}]{EmeryPRB1997}%
  \BibitemOpen
  \bibfield  {author} {\bibinfo {author} {\bibfnamefont {V.~J.}\ \bibnamefont
  {Emery}}, \bibinfo {author} {\bibfnamefont {S.~A.}\ \bibnamefont
  {Kivelson}},\ and\ \bibinfo {author} {\bibfnamefont {O.}~\bibnamefont
  {Zachar}},\ }\bibfield  {title} {\bibinfo {title} {Spin-gap proximity effect
  mechanism of high-temperature superconductivity},\ }\href
  {https://doi.org/10.1103/PhysRevB.56.6120} {\bibfield  {journal} {\bibinfo
  {journal} {Physical Review B}\ }\textbf {\bibinfo {volume} {56}},\ \bibinfo
  {pages} {6120} (\bibinfo {year} {1997})}\BibitemShut {NoStop}%
\bibitem [{\citenamefont {Kivelson}\ \emph {et~al.}(1998)\citenamefont
  {Kivelson}, \citenamefont {Fradkin},\ and\ \citenamefont
  {Emery}}]{KivelsonNature1998}%
  \BibitemOpen
  \bibfield  {author} {\bibinfo {author} {\bibfnamefont {S.~A.}\ \bibnamefont
  {Kivelson}}, \bibinfo {author} {\bibfnamefont {E.}~\bibnamefont {Fradkin}},\
  and\ \bibinfo {author} {\bibfnamefont {V.~J.}\ \bibnamefont {Emery}},\
  }\bibfield  {title} {\bibinfo {title} {Electronic liquid-crystal phases of a
  doped {M}ott insulator},\ }\href {https://doi.org/10.1038/31177} {\bibfield
  {journal} {\bibinfo  {journal} {Nature}\ }\textbf {\bibinfo {volume} {393}},\
  \bibinfo {pages} {550} (\bibinfo {year} {1998})}\BibitemShut {NoStop}%
\bibitem [{\citenamefont {Perkovi\'{c}}\ \emph {et~al.}(1995)\citenamefont
  {Perkovi\'{c}}, \citenamefont {Dahmen},\ and\ \citenamefont
  {Sethna}}]{PerkovicPRL1995}%
  \BibitemOpen
  \bibfield  {author} {\bibinfo {author} {\bibfnamefont {O.}~\bibnamefont
  {Perkovi\'{c}}}, \bibinfo {author} {\bibfnamefont {K.}~\bibnamefont
  {Dahmen}},\ and\ \bibinfo {author} {\bibfnamefont {J.}~\bibnamefont
  {Sethna}},\ }\bibfield  {title} {\bibinfo {title} {Avalanches, {B}arkhausen
  noise, and plain old criticality},\ }\href
  {https://doi.org/10.1103/PhysRevLett.75.4528} {\bibfield  {journal} {\bibinfo
   {journal} {Physical Review Letters}\ }\textbf {\bibinfo {volume} {75}},\
  \bibinfo {pages} {4528} (\bibinfo {year} {1995})}\BibitemShut {NoStop}%
\bibitem [{Note1()}]{Note1}%
  \BibitemOpen
  \bibinfo {note} {Here, $z$ is the dynamical critical exponent, which is of
  order 1.}\BibitemShut {Stop}%
\bibitem [{\citenamefont {Nie}\ \emph {et~al.}(2014)\citenamefont {Nie},
  \citenamefont {Tarjus},\ and\ \citenamefont {Kivelson}}]{NiePNAS2014}%
  \BibitemOpen
  \bibfield  {author} {\bibinfo {author} {\bibfnamefont {L.}~\bibnamefont
  {Nie}}, \bibinfo {author} {\bibfnamefont {G.}~\bibnamefont {Tarjus}},\ and\
  \bibinfo {author} {\bibfnamefont {S.~A.}\ \bibnamefont {Kivelson}},\
  }\bibfield  {title} {\bibinfo {title} {Quenched disorder and vestigial
  nematicity in the pseudogap regime of the cuprates},\ }\href
  {https://doi.org/10.1073/pnas.1406019111} {\bibfield  {journal} {\bibinfo
  {journal} {Proceedings of the National Academy of Sciences}\ }\textbf
  {\bibinfo {volume} {111}},\ \bibinfo {pages} {7980} (\bibinfo {year}
  {2014})}\BibitemShut {NoStop}%
\bibitem [{\citenamefont {Carlson}\ \emph {et~al.}(2006)\citenamefont
  {Carlson}, \citenamefont {Dahmen}, \citenamefont {Fradkin},\ and\
  \citenamefont {Kivelson}}]{CarlsonPRL2006}%
  \BibitemOpen
  \bibfield  {author} {\bibinfo {author} {\bibfnamefont {E.}~\bibnamefont
  {Carlson}}, \bibinfo {author} {\bibfnamefont {K.~A.}\ \bibnamefont {Dahmen}},
  \bibinfo {author} {\bibfnamefont {E.}~\bibnamefont {Fradkin}},\ and\ \bibinfo
  {author} {\bibfnamefont {S.}~\bibnamefont {Kivelson}},\ }\bibfield  {title}
  {\bibinfo {title} {Hysteresis and noise from electronic nematicity in
  high-temperature superconductors},\ }\href
  {https://doi.org/10.1103/PhysRevLett.96.097003} {\bibfield  {journal}
  {\bibinfo  {journal} {Physical Review Letters}\ }\textbf {\bibinfo {volume}
  {96}},\ \bibinfo {pages} {097003} (\bibinfo {year} {2006})}\BibitemShut
  {NoStop}%
\bibitem [{\citenamefont {Carlson}\ \emph {et~al.}(2015)\citenamefont
  {Carlson}, \citenamefont {Liu}, \citenamefont {Phillabaum},\ and\
  \citenamefont {Dahmen}}]{CarlsonJSNM2015}%
  \BibitemOpen
  \bibfield  {author} {\bibinfo {author} {\bibfnamefont {E.~W.}\ \bibnamefont
  {Carlson}}, \bibinfo {author} {\bibfnamefont {S.}~\bibnamefont {Liu}},
  \bibinfo {author} {\bibfnamefont {B.}~\bibnamefont {Phillabaum}},\ and\
  \bibinfo {author} {\bibfnamefont {K.~A.}\ \bibnamefont {Dahmen}},\ }\bibfield
   {title} {\bibinfo {title} {Decoding spatial complexity in strongly
  correlated electronic systems},\ }\href
  {https://doi.org/10.1007/s10948-014-2898-0} {\bibfield  {journal} {\bibinfo
  {journal} {Journal of Superconductivity and Novel Magnetism}\ }\textbf
  {\bibinfo {volume} {28}},\ \bibinfo {pages} {1237–1243} (\bibinfo {year}
  {2015})}\BibitemShut {NoStop}%
\bibitem [{Note2()}]{Note2}%
  \BibitemOpen
  \bibinfo {note} {Quenched disorder can also introduce randomness in the
  couplings $J$, also known as random bond disorder. In the presence of both
  random bond and random field disorder, the critical behavior is controlled by
  the random field fixed point.}\BibitemShut {Stop}%
\bibitem [{\citenamefont {Elias}\ \emph {et~al.}(1956)\citenamefont {Elias},
  \citenamefont {Feinstein},\ and\ \citenamefont {Shannon}}]{1056816}%
  \BibitemOpen
  \bibfield  {author} {\bibinfo {author} {\bibfnamefont {P.}~\bibnamefont
  {Elias}}, \bibinfo {author} {\bibfnamefont {A.}~\bibnamefont {Feinstein}},\
  and\ \bibinfo {author} {\bibfnamefont {C.}~\bibnamefont {Shannon}},\
  }\bibfield  {title} {\bibinfo {title} {A note on the maximum flow through a
  network},\ }\href {https://doi.org/10.1109/TIT.1956.1056816} {\bibfield
  {journal} {\bibinfo  {journal} {IRE Transactions on Information Theory}\
  }\textbf {\bibinfo {volume} {2}},\ \bibinfo {pages} {117} (\bibinfo {year}
  {1956})}\BibitemShut {NoStop}%
\bibitem [{\citenamefont {Goldberg}(2009)}]{GoldbergAndrewVTPAf}%
  \BibitemOpen
  \bibfield  {author} {\bibinfo {author} {\bibfnamefont {A.~V.}\ \bibnamefont
  {Goldberg}},\ }\bibfield  {title} {{\selectlanguage {english}\bibinfo {title}
  {Two-level push-relabel algorithm for the maximum flow problem}},\ }in\
  \href@noop {} {{\selectlanguage {english}\emph {\bibinfo {booktitle}
  {Algorithmic Aspects in Information and Management}}}},\ \bibinfo {series and
  number} {Lecture Notes in Computer Science}\ (\bibinfo  {publisher} {Springer
  Berlin Heidelberg},\ \bibinfo {address} {Berlin, Heidelberg},\ \bibinfo
  {year} {2009})\ pp.\ \bibinfo {pages} {212--225}\BibitemShut {NoStop}%
\bibitem [{\citenamefont {Picard}\ and\ \citenamefont
  {Ratliff}(1975)}]{PicardJ.C1975Mcar}%
  \BibitemOpen
  \bibfield  {author} {\bibinfo {author} {\bibfnamefont {J.~C.}\ \bibnamefont
  {Picard}}\ and\ \bibinfo {author} {\bibfnamefont {H.~D.}\ \bibnamefont
  {Ratliff}},\ }\bibfield  {title} {{\selectlanguage {english}\bibinfo {title}
  {Minimum cuts and related problems}},\ }\href@noop {} {\bibfield  {journal}
  {\bibinfo  {journal} {Networks}\ }\textbf {\bibinfo {volume} {5}},\ \bibinfo
  {pages} {357} (\bibinfo {year} {1975})}\BibitemShut {NoStop}%
\bibitem [{\citenamefont {Hacker}\ \emph {et~al.}(2014)\citenamefont {Hacker},
  \citenamefont {Yang},\ and\ \citenamefont {McCartney}}]{rcac-purdue}%
  \BibitemOpen
  \bibfield  {author} {\bibinfo {author} {\bibfnamefont {T.}~\bibnamefont
  {Hacker}}, \bibinfo {author} {\bibfnamefont {B.}~\bibnamefont {Yang}},\ and\
  \bibinfo {author} {\bibfnamefont {G.}~\bibnamefont {McCartney}},\ }\bibfield
  {title} {\bibinfo {title} {{Empowering Faculty: A Campus Cyberinfrastructure
  Strategy for Research Communities}},\ }\href
  {https://er.educause.edu/articles/2014/7/empowering-faculty-a-campus-cyberinfrastructure-strategy-for-research-communities}
  {\bibfield  {journal} {\bibinfo  {journal} {Educause Review}\ } (\bibinfo
  {year} {2014})}\BibitemShut {NoStop}%
\bibitem [{\citenamefont {Stauffer}(1979)}]{StaufferPhysRep1979}%
  \BibitemOpen
  \bibfield  {author} {\bibinfo {author} {\bibfnamefont {D.}~\bibnamefont
  {Stauffer}},\ }\bibfield  {title} {\bibinfo {title} {Scaling theory of
  percolation clusters},\ }\href {https://doi.org/10.1016/0370-1573(79)90060-7}
  {\bibfield  {journal} {\bibinfo  {journal} {Physics Reports}\ }\textbf
  {\bibinfo {volume} {54}},\ \bibinfo {pages} {1} (\bibinfo {year}
  {1979})}\BibitemShut {NoStop}%
\bibitem [{\citenamefont {Janke}\ and\ \citenamefont
  {Schakel}(2005)}]{JankePRE2005}%
  \BibitemOpen
  \bibfield  {author} {\bibinfo {author} {\bibfnamefont {W.}~\bibnamefont
  {Janke}}\ and\ \bibinfo {author} {\bibfnamefont {A.~M.~J.}\ \bibnamefont
  {Schakel}},\ }\bibfield  {title} {\bibinfo {title} {Fractal structure of spin
  clusters and domain walls in the two-dimensional {I}sing model},\ }\href
  {https://doi.org/10.1103/PhysRevE.71.036703} {\bibfield  {journal} {\bibinfo
  {journal} {Physical Review E}\ }\textbf {\bibinfo {volume} {71}},\ \bibinfo
  {pages} {036703} (\bibinfo {year} {2005})}\BibitemShut {NoStop}%
\bibitem [{\citenamefont {Saberi}\ and\ \citenamefont
  {Dashti-Naserabadi}(2010)}]{SaberiEPL2010}%
  \BibitemOpen
  \bibfield  {author} {\bibinfo {author} {\bibfnamefont {A.~A.}\ \bibnamefont
  {Saberi}}\ and\ \bibinfo {author} {\bibfnamefont {H.}~\bibnamefont
  {Dashti-Naserabadi}},\ }\bibfield  {title} {\bibinfo {title} {Three
  dimensional {I}sing model, percolation theory and conformal invariance},\
  }\href {https://doi.org/10.1209/0295-5075/92/67005} {\bibfield  {journal}
  {\bibinfo  {journal} {Europhysics Letters}\ }\textbf {\bibinfo {volume}
  {92}},\ \bibinfo {pages} {67005} (\bibinfo {year} {2010})}\BibitemShut
  {NoStop}%
\bibitem [{\citenamefont {Livet}(1991)}]{Livet_1991}%
  \BibitemOpen
  \bibfield  {author} {\bibinfo {author} {\bibfnamefont {F.}~\bibnamefont
  {Livet}},\ }\bibfield  {title} {\bibinfo {title} {The cluster updating monte
  carlo algorithm applied to the 3 d ising problem},\ }\href
  {https://doi.org/10.1209/0295-5075/16/2/003} {\bibfield  {journal} {\bibinfo
  {journal} {Europhysics Letters ({EPL})}\ }\textbf {\bibinfo {volume} {16}},\
  \bibinfo {pages} {139} (\bibinfo {year} {1991})}\BibitemShut {NoStop}%
\bibitem [{\citenamefont {Talapov}\ and\ \citenamefont
  {Blöte}(1996)}]{Talapov_1996}%
  \BibitemOpen
  \bibfield  {author} {\bibinfo {author} {\bibfnamefont {A.~L.}\ \bibnamefont
  {Talapov}}\ and\ \bibinfo {author} {\bibfnamefont {H.~W.~J.}\ \bibnamefont
  {Blöte}},\ }\bibfield  {title} {\bibinfo {title} {The magnetization of the
  3d ising model},\ }\href {https://doi.org/10.1088/0305-4470/29/17/042}
  {\bibfield  {journal} {\bibinfo  {journal} {Journal of Physics A:
  Mathematical and General}\ }\textbf {\bibinfo {volume} {29}},\ \bibinfo
  {pages} {5727} (\bibinfo {year} {1996})}\BibitemShut {NoStop}%
\end{thebibliography}%

\clearpage
\onecolumngrid
\setcounter{equation}{0}
\setcounter{figure}{0}
\setcounter{table}{0}
\setcounter{section}{0}
\setcounter{page}{1}
\renewcommand{\theequation}{S\arabic{equation}}
\renewcommand{\thefigure}{S\arabic{figure}}
\renewcommand{\thetable}{S\Roman{table}}
\renewcommand{\thesection}{S\arabic{section}}

\begin{center}
\textbf{\large Supplementary Material for\\
\bigskip
\mytitle}
\end{center}


%

\subsection{Setting the gaussian window of the FT}
\begin{figure}
\center
 \includegraphics[width=.5\columnwidth]{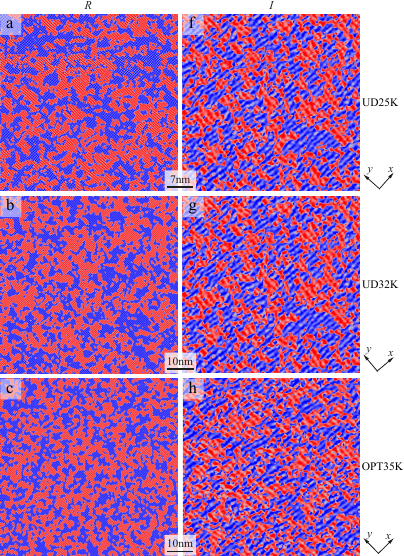}
  \includegraphics[width=.5\columnwidth]{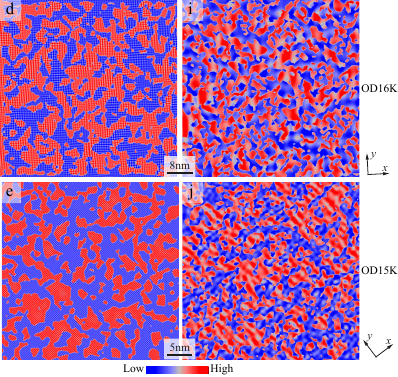}
\caption{\textbf{Ising domains.} (\textbf{a-j}) Mapping Ising domains from $R$ (\textbf{a-e}) and $I$ maps (\textbf{f-j}) in various samples, colored red ($\sigma=1$, Q$_{x}^{**}$/Q$_{x}^{*}$ dominates) and blue ($\sigma=-1$, Q$_{y}^{**}$/Q$_{y}^{*}$ dominates) to indicate the local unidirectional orientations. We map the Ising domains in (\textbf{f-j}) by comparing the integrated FT intensity between around $\vec{q}=\pm Q_x^{*}$ and $\vec{q}=\pm Q_y^{*}$ using a similar round integration window centered at $\pm Q_x^{*}$ and $\pm Q_y^{*}$ (dashed circles in Supplementary Fig.\ S1b). The Gaussian width $L=0.9a_0$ has been chosen to calculate the Ising maps from $I$ map datasets (\textbf{f-j}). For clarity, all maps have been fourier-filtered to include only the FT power spectral density surrounding $\textrm{Q}^{**}$ (\textbf{a-e}) or $\textrm{Q}^{*}$ (\textbf{f-j}) peaks.
}
  \label{fig:Ising-Maps}
\end{figure}

In applying Eqn.~\ref{eqn:FT} to the data, the width of the gaussian window of the FT is set by the parameter $L$.
In order to find the optimal $L$, we define a quality factor $H= \textrm{F}_x+\textrm{F}_{y}^{'}-\textrm{F}_y-\textrm{F}_{x}^{'}$. Here $\textrm{F}_x$ and $\textrm{F}_y$ represent the integrated FT intensity around $\pm Q_x^{**}$ and $\pm Q_y^{**}$ of the red-colored regions of raw $R$ map ($\sigma = +1$, Fig.~\ref{fig:UD32K-mapping}(b)), while $\textrm{F}_x^{'}$ and $\textrm{F}_y^{'}$ the integrated FT intensity around $\pm Q_x^{**}$ and $\pm Q_y^{**}$ of the blue-colored regions of raw $R$ map ($\sigma =- 1$, Fig.~\ref{fig:UD32K-mapping}(c)). A larger $H$ value means a higher quality of the Ising map. We therefore choose the optimal $L$ and $f$  by maximizing $H$. In our study, we find that $L$ is typically in the range of (0.4-0.9)$a_0$. For consistency, we have chosen $L=0.6a_{0}$ for all $R$ map datesets in Fig.~\ref{fig:UD32K-mapping}(a) and Supplementary Fig.~\ref{fig:Ising-Maps}.

\subsection{Critical Points of the Models}
At a critical point, a system displays critical, power law behavior at all length scales.  
Near but not at a critical point, a system displays critical behavior below a finite correlation length $\xi$, which approaches infinity at the critical point.
The critical fixed points which control the continuous phase transitions contained in the models of Eqn.~\ref{eqn:model} are as follows:  With no quenched disorder ($\Delta = 0$), the universality class is that of the clean Ising model.  The critical exponents are different in two dimensions (at the C-2D fixed point), {\em vs.} three dimensions (at the C-3D fixed point).   In the noninteracting limit $J_{||} = J_{\perp} = 0$, Eqn.~\ref{eqn:model} describes the percolation model, in which the stripe orientation at each site $\sigma_i = \pm 1$ takes a value independently of its neighbor.  The only time this model can display criticality when viewed on a two-dimensional FOV is when the probability of having, say, $\sigma_i = +1$ on each site is near $p = 0.59$, (or its complement $p = 1-0.59$) {\em i.e.} at P-2D.   When both interactions and random field disorder are present, the universality class is that of the three-dimensional random field Ising model (RF-3D) when $J_{\perp} > 0$, or that of RF-2D if $J_{\perp} = 0$. 
Any finite coupling $J^{\perp} > 0$ in the third dimension is relevant in the renormalization group sense, even in a strongly layered model with $J^{\perp} \ll J^{||}$, 
so that ultimately the critical behavior is that of the 3D model at long enough length scales, whether clean (C-3D) or random field (RF-3D).

\subsection{Comparing data-derived critical exponents to theoretical models}

As evident in the data, the nematic orientation-orientation correlation function
(spin-spin correlation function in the Ising language) does not display robust power law behavior.  On the other hand, the connectivity function
(defined as the probability that two aligned sites are connected by the same cluster) does show robust power law behavior.  Because this behavior is known to be qualitatively consistent with uncorrelated percolation fixed points,\cite{StaufferPhysRep1979} one might surmise that the intricate pattern formation observed on the surface of this set of materials is simply due to uncorrelated percolation, which would lead to the dubious conclusion that the role of {\em both} interactions and disorder is {\em irrelevant} in the renormalization group sense for the multiscale pattern formation in this material. While the extracted exponent for the anomalous dimension of the connectivity function $d-2+\eta_{\mathrm{connect}}$ is somewhat close to that of uncorrelated 2D percolation, the hull fractal dimension in that model is $d_h = 7/4=1.75$,\cite{JankePRE2005} which is contradicted by the experimental data shown here.  Note that 3D percolation at criticality is also ruled out, not by the large discrepancy between the theoretical and data-derived values of  the exponent $d-2+\eta_{\mathrm{connect}}$, but rather by the fact that the clusters do not display fractal properties at a free surface at the 3D percolation fixed point.  This is underscored by the fact that the 
connectivity function in that model is power law in the bulk, but is not power law on a 2D slice. 
Therefore, the pattern formation in these Bi2201 samples is {\em not} controlled by uncorrelated percolation, but there must exist some other fixed point(s) at which the connectivity function is power law.

Our Monte Carlo simulations of the clean 2D Ising model at the critical temperature reveal that in fact, the connectivity function is also power law at the clean 2D fixed point, with an exponent $d-2+\eta_{\mathrm{connect}} = 0.098 \pm .005$, which also serves to rule out the clean 2D fixed point as the origin of this behavior in the data.\cite{C-3Dx} 
Note that the exponents associated with the geometric clusters at the C-2D fixed point are $\beta_p/\nu_p= 5/96$ and $d_{v,p}=187/96$,\cite{JankePRE2005}, where the volume fractal dimension $d_{v,p}$ is derived from the fractal structure of the geometric clusters as they percolate at $T_p = T_c$. Inserting these values into the exponent relation $2 \beta_p/\nu_p=2(d-d_{v,p})$ yields $2*(5/96)=2*(2-187/96)$, indicating that the geometric clusters also satisfy scaling relations at C-2D.

\begin{figure}
    \centering
    \includegraphics{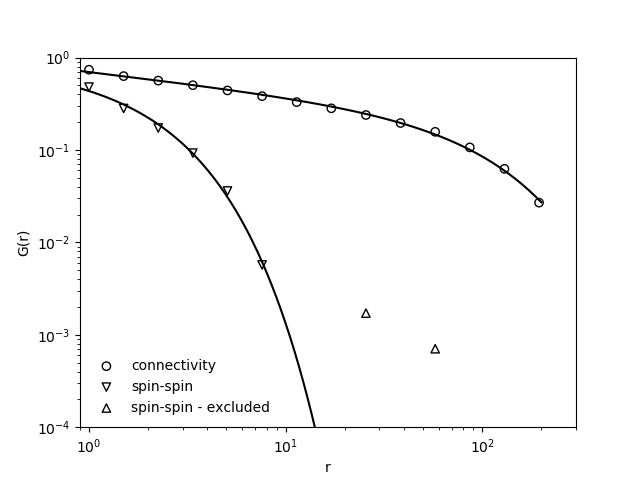}
    \caption{An example of the pair connectivity function and the spin-spin correlation function for a 256x256 window on a free surface of a 512x512x512 3D RFIM system with open boundary conditions on the top and bottom surface and periodic boundary conditions in all other directions. Here, $T=0$ and the disorder strength $\Delta = 3 J$ (see Eqn.~\ref{eqn:model} in the main text).
    The open circles represent the pair connectivity function $G_{\rm conn}(r)$, and the open triangles represent the spin-spin correlation function, $G_{\rm spin}(r)$.  The solid lines are fits to a power law times an exponential, $G(r) \propto e^{-r/\xi}/r^{d-2+\eta}$.
    Points beyond $r = 11$ sites have been excluded from the fit of the spin-spin correlation function, 
    because the value has effectively become zero beyond that point, with many values being negative, but all bounded by $|G_{\rm spin}(r)| \lesssim .002$.     
    Note that while the spin-spin correlation function is not robustly power law, the pair connectivity function is. Here, $\xi_{\rm spin} = 1.59 \pm 0.22$ while $\xi_{\rm pair} = 103.1 \pm 4.3$ .}
    \label{fig:3drfim-corrfxns}
\end{figure}

\begin{figure}
    \centering
    \includegraphics{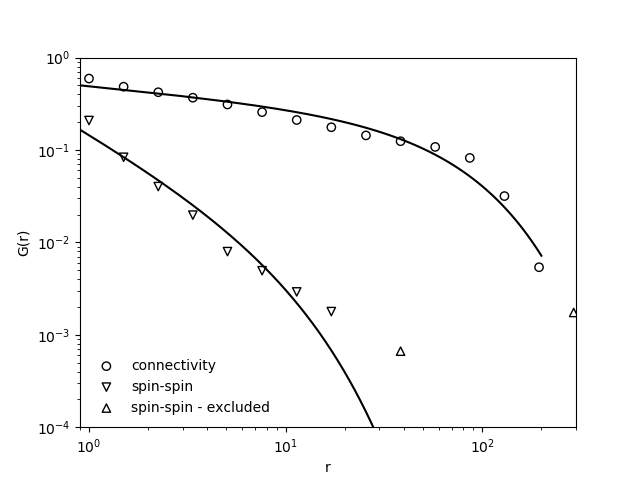}
    \caption{An example of the pair connectivity function and the spin-spin correlation function for a 256x256 window on a free surface of a 840x840x840 3D clean Ising model with open boundary conditions on the top and bottom surface and periodic boundary conditions in all other directions. Here, $T=4.51 J \lesssim T_c = 4.51152786 J$.\cite{Livet_1991,Talapov_1996} The open circles represent the pair connectivity function $G_{\rm conn}(r)$, and the open triangles represent the spin-spin correlation function, $G_{\rm spin}(r)$.  The solid lines are fits to a power law times an exponential, $G(r) \propto e^{-r/\xi}/r^{d-2+\eta}$.
    Points beyond $r = 25$ sites have been excluded from the fit of the spin-spin correlation function, 
    because the value has effectively become zero beyond that point, with many values being negative, but all bounded by $|G_{\rm spin}(r)| \lesssim .0005$.     
    Note that while the spin-spin correlation function is not robustly power law, the pair connectivity function is. Here, $\xi_{\rm spin} = 8.2 \pm 3.9$ while $\xi_{\rm pair} = 62.7 \pm 8.3 $.  The fact that the spin-spin correlation function shows less than a decade of scaling even within $0.03\%$ of $T_c$ is indicative of how narrow the critical region is for C-3D. 
    }
    \label{fig:C-3Ds-corrfxns}
\end{figure}

The behavior in question concerns the observed correlation functions on a free surface of the material, rather than the bulk correlation functions. Our simulations of both the clean Ising model and the random  field Ising model in three dimensions reveal that in fact, the connectivity function {\em at a free surface} displays power law behavior at the thermodynamic critical point.
(See Figs.\ref{fig:3drfim-corrfxns} and ~\ref{fig:C-3Ds-corrfxns}.)
For the clean model, this behavior is directly related to the fact that while 3D geometric clusters do not undergo a (correlated) percolation point until $T_p < T_c$, geometric clusters on a 2D slice undergo a (correlated) percolation point right at $T_p^{\rm 2D~slice} = T_c$,\cite{SaberiEPL2010,C-3Dx} indicating that the geometric clusters on the surface also undergo a (correlated) percolation point at the same temperature.  
Our results indicate that in the same way, geometric clusters on the free surface of a 3D random field model also undergo a (correlated) percolation point at $T_p^{\rm 2D~slice}$, leading to the robust power law connectivity function revealed in our simulations of the 3D random field Ising model. Our results for the critical exponents at a free surface of 
the the clean and random field models in three dimensions are shown in Fig.~\ref{fig:exponent-comparison}(c). As can be seen from the figure, the data match quite well with either a clean or random field  3D Ising model for this measure.


We have also calculated the connectivity function of the 2D random field Ising model. For weak random field strength, we find that the connectivity function is also a power law in this model, with exponent $d-2+\eta_{\rm connect} \approx 0.29 \pm .01$.

The pair connectivity function and the spin-spin correlation function are plotted in Fig.~\ref{fig:3drfim-corrfxns} for one disorder configuration, on a 256x256 window on the open surface of the three-dimensional random field Ising model of system size 512x512x512  with open boundary conditions in the z direction and periodic boundary conditions in the x and y directions. Both plots are logarithmically binned and fit to a function of the form $G(r) = C e^{-r/\xi} r^{-(d-2+\eta)}$.  In the spin-spin correlation plot, for several of the computed points starting with $r \approx 11.4$, the measured value of $G(r)$ is negative. These negative points cannot be displayed on a log scale and cannot be included in the exponentially decaying power law fit.  The downward trend in $G(r)$ followed by values fluctuating around 0 indicate that the error in the measured values due to finite size effects is of similar size to the values themselves. Thus, all values after $r=11$ are excluded from the fit. Note that while the spin-spin correlation function is not robustly power law, the pair connectivity function is. The same behavior appears in the experimental data as shown Fig.~\ref{fig:power-law-fits}(d).  While it is well-known that the pair connectivity function is power law near criticality in uncorrelated percolation, it was not previously known that the pair connectivity function (defined on a surface) is power law near criticality of the 3D RFIM.  

The pair connectivity and spin-spin correlation function are plotted in Fig. ~\ref{fig:C-3Ds-corrfxns} for a 256x256 window on a free surface of a clean 3D Ising model of system size 840x840x840 with open boundary conditions in the z direction and periodic boundary conditions in the x and y directions simulated at $T=T_c$. Both plots are logarithmically binned and fit to a function of the form $G(r) = C e^{-r/\xi} r^{-(d-2+\eta)}$.  In the spin-spin correlation plot, for several of the computed points starting with $r \approx 38.4$, the measured value of $G(r)$ is negative. Because these points cannot be included in te fit, all subsequent points have also been omitted. While the spin-spin correlation function is not robustly power law, the pair connectivity is.


\begin{figure}
\center
 \includegraphics[width=1\columnwidth]{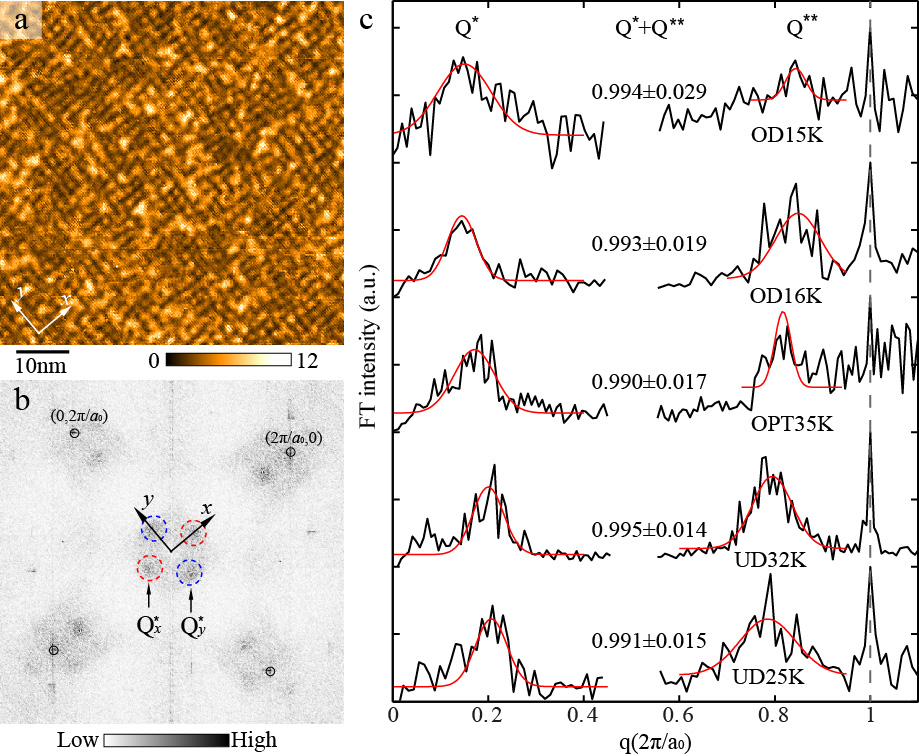}
\caption{\textbf{The sum of wavevector Q$^{*}$ + Q$^{**}$ $\simeq$ 1.} (\textbf{a}) Map of $\textrm{d}I/d\textrm{V}$ integrated over the energy range of -20 mV and 20 mV (hereafter denoted as $I$ map), acquired in the same field of view shown in Fig.\ 2a. Such map highlights Q$^{*}$ modulations. The brighter color corresponds to larger $I$. The $\sim4a_{0}$  charge modulation is quite prominent in the real space. The conductance $\textrm{d}I/d\textrm{V}$ maps are acquired at $I$ = 400 pA and $V_s$= -200 mV. (\textbf{b}) Fourier transform of $I$ map in (\textbf{a}), with Bragg vectors ($\pm$1, 0)$2\pi/a_{0}$ and (0, $\pm$1)$2\pi/a_{0}$ marked by black circles. The wavevectors Q$_{x}^{*}$ $\sim$ (1/4, 0)$2\pi/a_{0}$ and Q$_{y}^{*}$ $\sim$ (0, 1/4)$2\pi/a_{0}$ from the charge modulation are pronounced and marked by dashed circles and arrows, from which Q$^{*}$ wavevectors are extracted. The central broad FT intensities have been eliminated by subtracting the radial background, obtained by a power-law fit of the FT intensities near the diagonal $x\pm y$ directions. (\textbf{c}) Gaussian fits (red curves) of FT intensities in various samples. The left and right black curves are extracted from the fourier-transformed images of $I$ and $R$ maps, respectively. Note that Q$^{*}$ + Q$^{**}$ $\simeq$ 1 holds for all samples. The errors
indicate the standard derivation of Q$^{*}$ + Q$^{**}$ values obtained by choosing different wavevector windows for the Gaussian fits.
}
  \label{fig:lineshapes}
\end{figure}

\newpage 

\begin{figure}
\center
 \includegraphics[width=1\columnwidth]{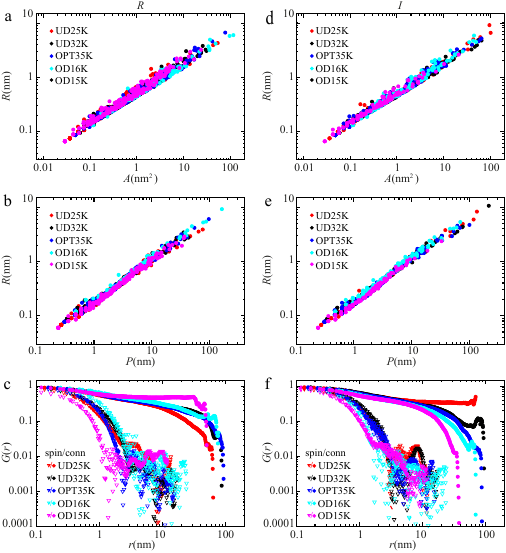}
\caption{\textbf{Cluster structure and correlation functions.} Statistics and correlation functions of Ising cluster derived from $R$ maps (\textbf{a-d}) and $I$ maps (\textbf{e-h}) in various samples. The empty triangles and solid circles in (\textbf{c}) and (\textbf{f}) indicate the spatial spin-spin and connectivity correlation functions, respectively.
}
  \label{fig:analysis-exponents}
\end{figure}

\begin{figure}
\center
  \includegraphics[width=.8\columnwidth]{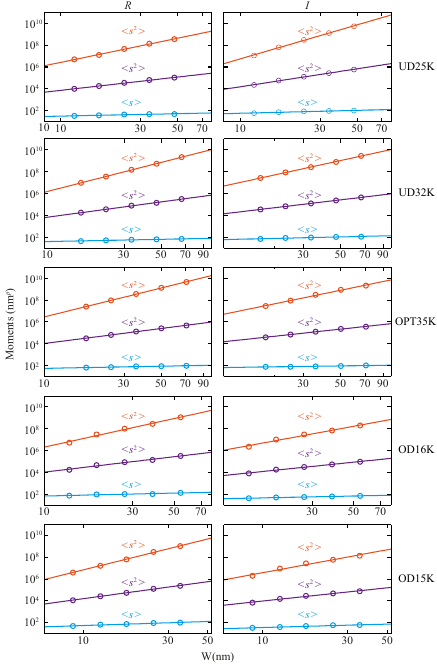}
\caption{
\textbf{Finite-size-scaling of moments for cluster size distribution.} A power law is generally found between each moment and crop size $W$ for all Ising maps in Supplementary Fig.~\ref{fig:Ising-Maps}, either from $R$ (\textbf{a-e}) and $I$ maps (\textbf{f-j}).
}
  \label{fig:cluster-moments}
\end{figure}
\begin{figure}
\begin{tabular}{cc}
    \includegraphics[width=65mm]{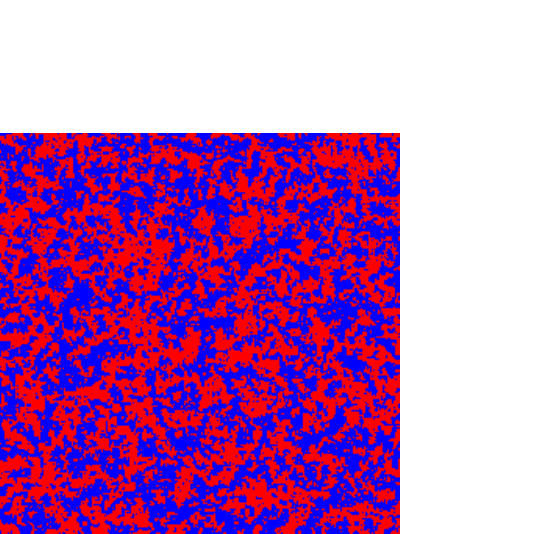} & \includegraphics[width=65mm]{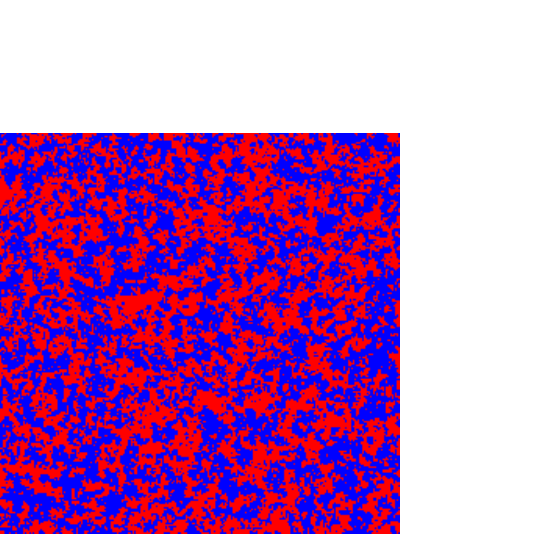} \\
    \includegraphics[width=65mm]{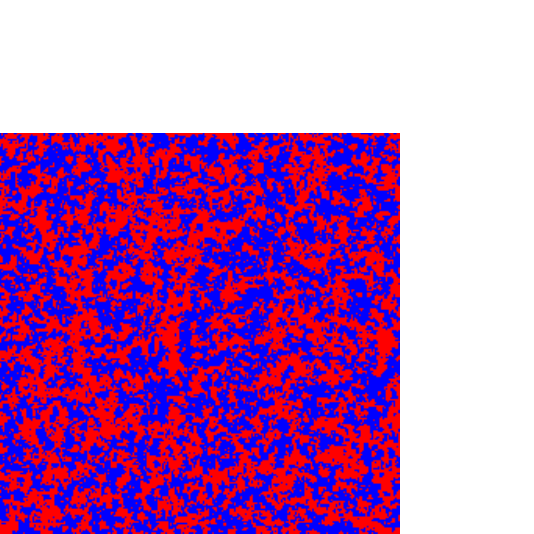} & \includegraphics[width=65mm]{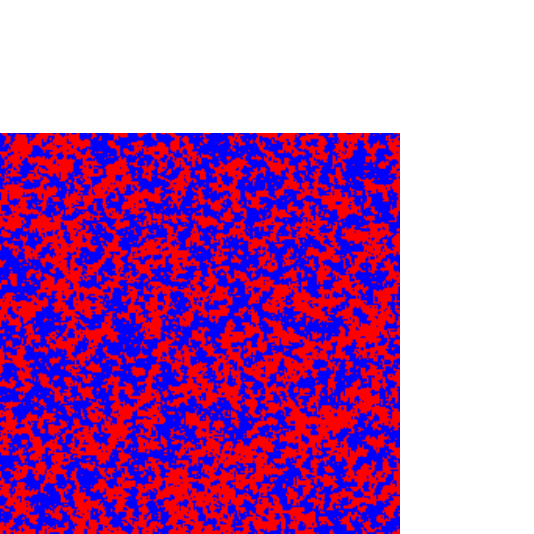} \\
    \includegraphics[width=65mm]{supplement/Images/3DRFIMImages/spin-3_0_1_0_0-surface.pdf} & \includegraphics[width=65mm]{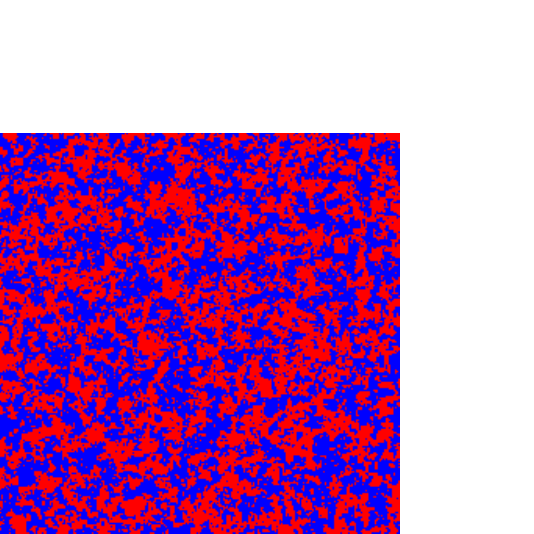} \\
\end{tabular}
\caption{Representative ground state configurations of stripe orientations on a free surface of the near-critical 3D RFIM at disorder strength $R=3 J$.  The images show windows of size 256$\times$256 on the free surface taken from exact calculations of the ground state of systems of size 512$\times$512$\times$512.
}
\end{figure}
\begin{figure}
\begin{tabular}{cc}
    \includegraphics[width=65mm]{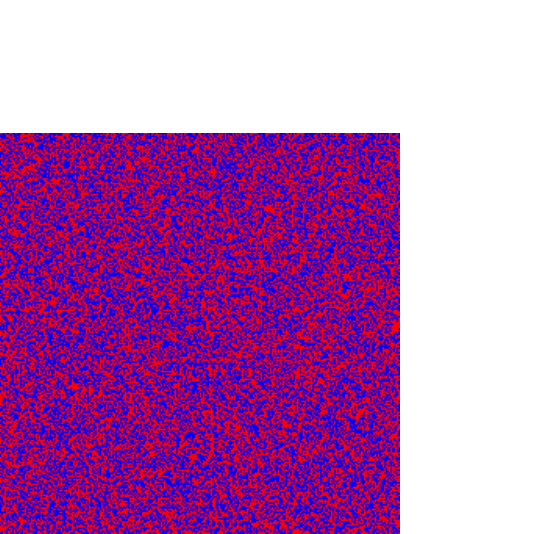} & \includegraphics[width=65mm]{supplement/Images/PercolationImages/spin-PercolationTest-256-0_0.591-0.pdf} \\
    \includegraphics[width=65mm]{supplement/Images/PercolationImages/spin-PercolationTest-256-0_0.591-0.pdf} & \includegraphics[width=65mm]{supplement/Images/PercolationImages/spin-PercolationTest-256-0_0.591-0.pdf} \\
    \includegraphics[width=65mm]{supplement/Images/PercolationImages/spin-PercolationTest-256-0_0.591-0.pdf} & \includegraphics[width=65mm]{supplement/Images/PercolationImages/spin-PercolationTest-256-0_0.591-0.pdf} \\
\end{tabular}
\caption{256x256 percolation images at $p=p_c$}
\end{figure}
\begin{figure}
\begin{tabular}{cc}
    \includegraphics[width=65mm]{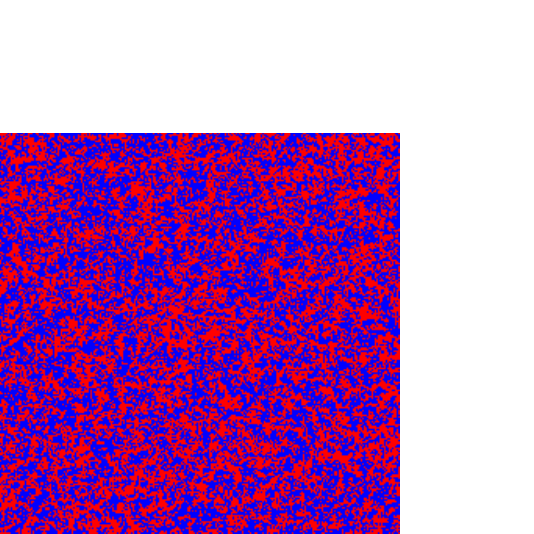} & \includegraphics[width=65mm]{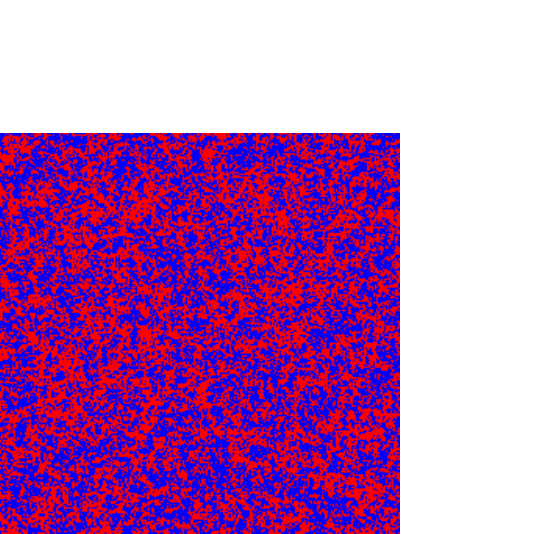} \\
    \includegraphics[width=65mm]{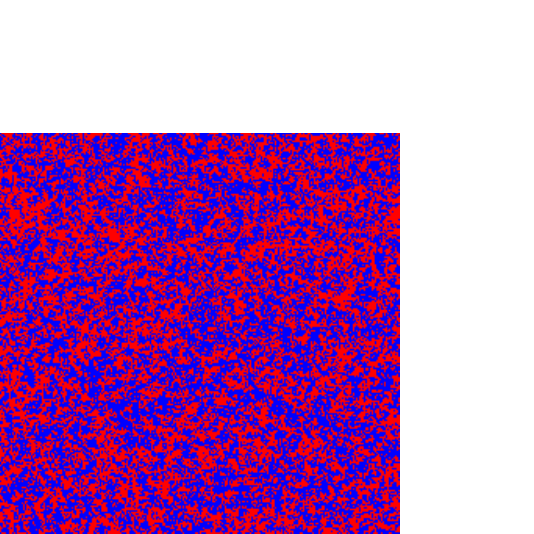} & \includegraphics[width=65mm]{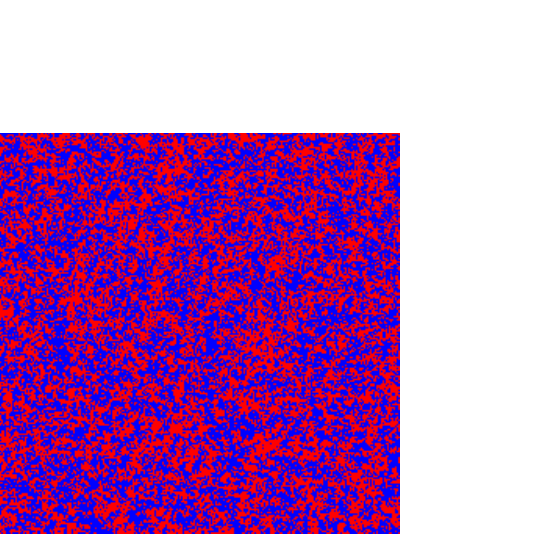} \\
    \includegraphics[width=65mm]{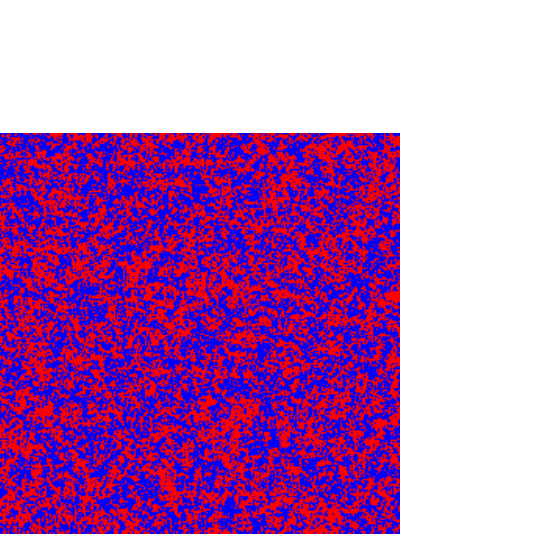} & \includegraphics[width=65mm]{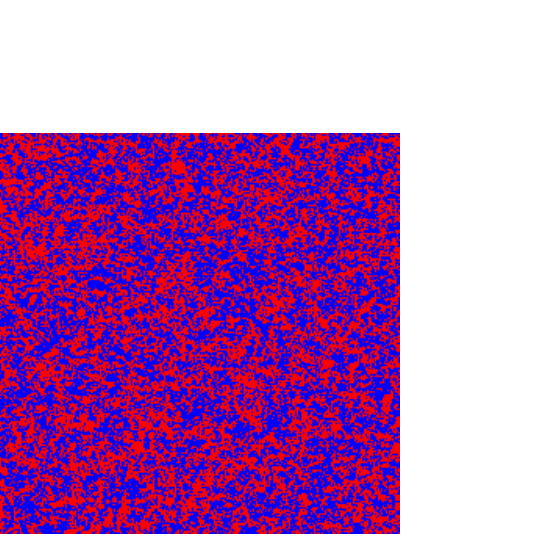} \\
\end{tabular}
\caption{256x256 windows of the surface of an 840x840x840 3D clean Ising model at$T=T_c$}
\end{figure}
\begin{figure}
\begin{tabular}{cc}
    \includegraphics[width=65mm]{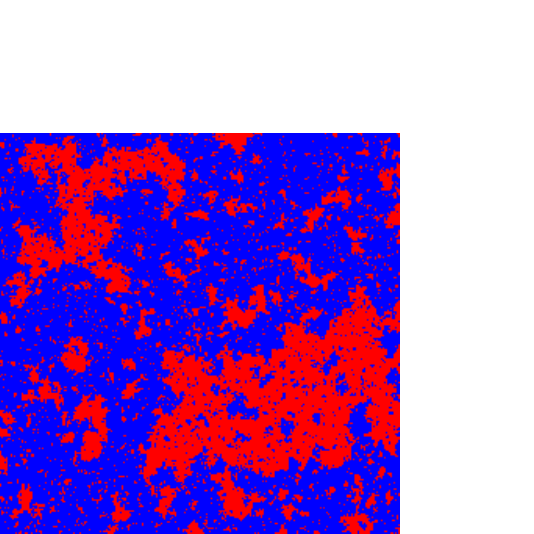} & \includegraphics[width=65mm]{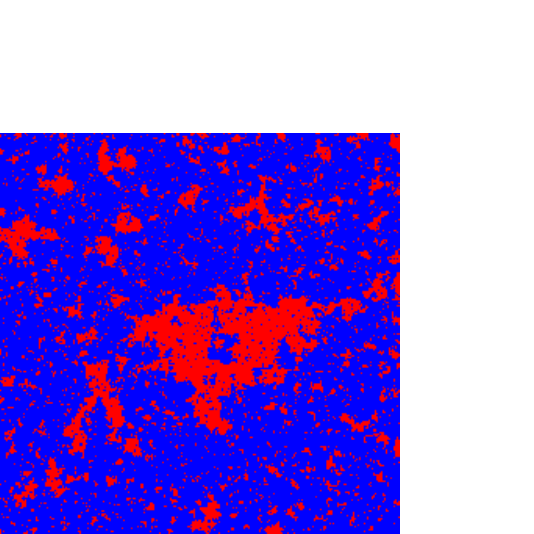} \\
    \includegraphics[width=65mm]{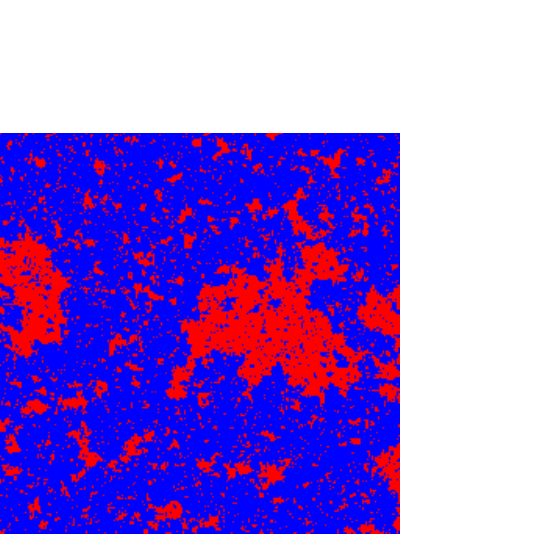} & \includegraphics[width=65mm]{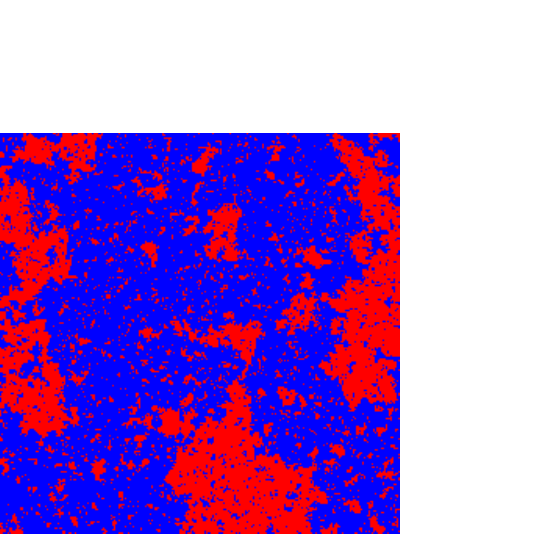} \\    \includegraphics[width=65mm]{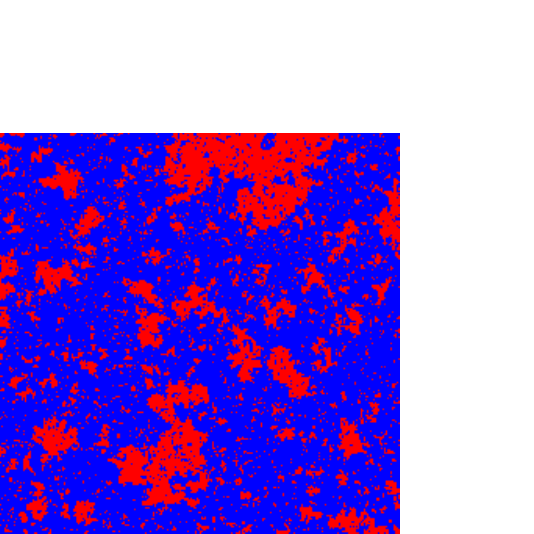} & \includegraphics[width=65mm]{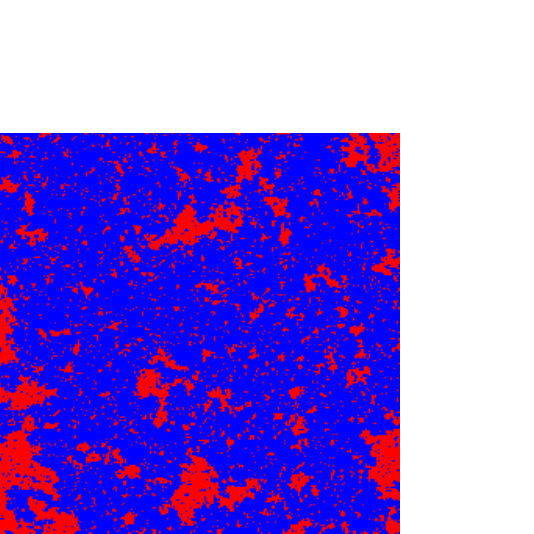} \\
\end{tabular}
\caption{256x256 windows of a 1000x1000 2D clean Ising model at $T=T_c$}
\end{figure}
\begin{figure}
\begin{tabular}{cc}
    \includegraphics[width=65mm]{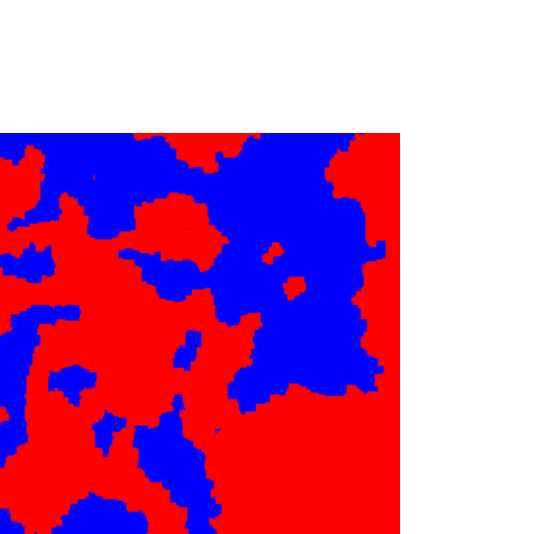} & \includegraphics[width=65mm]{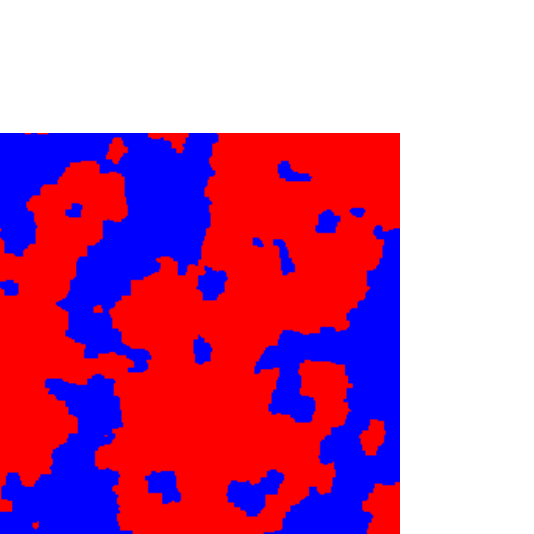} \\
    \includegraphics[width=65mm]{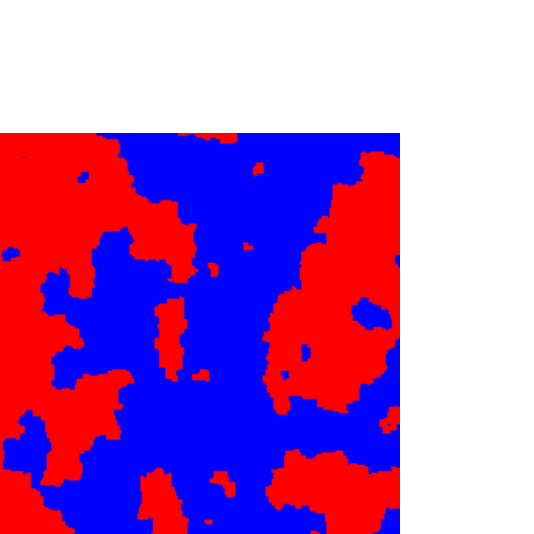} & \includegraphics[width=65mm]{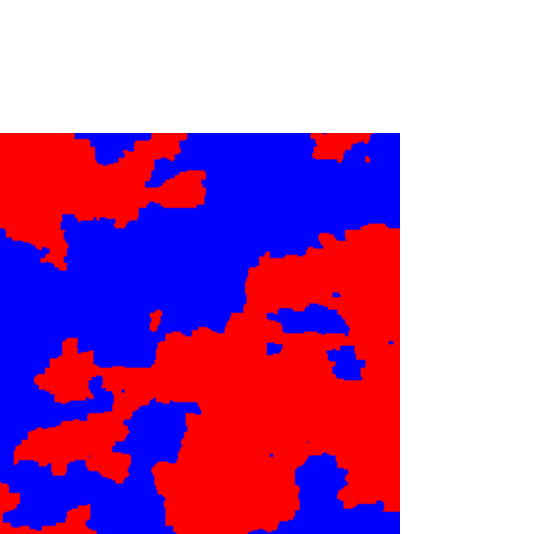} \\
    \includegraphics[width=65mm]{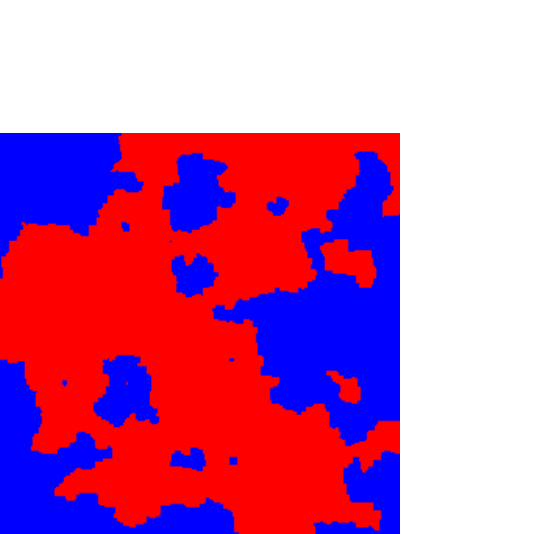} & \includegraphics[width=65mm]{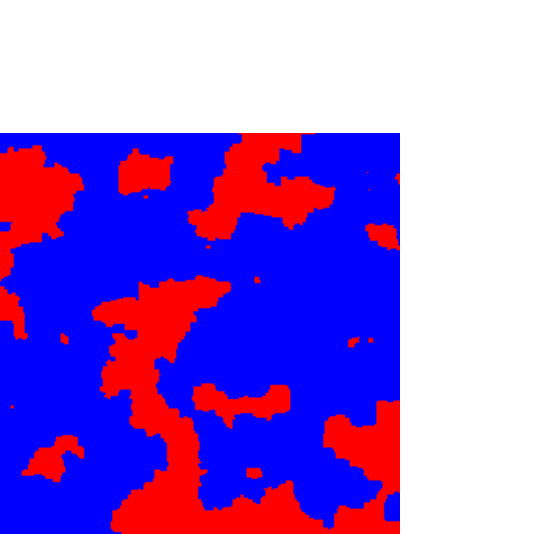} \\
\end{tabular}
    \caption{256x256 windows of a 512x512 2D random field Ising model at $T=0$ and $R=1$. Simulations run at $R=0.5$ and $R=0.6$ produced completely magnetized configurations for this size system.}
\end{figure}

\end{document}